\documentclass[10pt, twocolumn, comsoc]{IEEEtran}

\usepackage{graphicx,epsfig}
\usepackage[noadjust]{cite}
\usepackage{mcite}
\usepackage{amsfonts,helvet}
\usepackage{fancyhdr}
\usepackage{threeparttable}
\usepackage{epsf,epsfig}
\usepackage{amsthm}
\usepackage{amsmath}
\usepackage{siunitx}
\usepackage{amssymb}
\usepackage{dsfont}
\usepackage{subfigure}
\usepackage{color}
\usepackage[linesnumbered,ruled]{algorithm2e}
\usepackage{enumerate}
\usepackage{cancel}
\usepackage{graphicx}
\usepackage{comment}
\usepackage{siunitx}
\usepackage[colorlinks=true, linkcolor=blue]{hyperref}
\usepackage{xcolor}
\usepackage{mdframed}
\usepackage{multirow}

\newtheorem{remark}{Remark}

\SetKwInput{KwInput}{Initialize}
\SetKwInput{KwOutput}{Output}
\SetKwProg{Fn}{Stage 1}{}{}
\SetKwProg{Fn}{Stage 2}{}{}

\usepackage{eucal}

\begin{document}

\title{Energy-Efficient State Estimation with 1-Bit Sensing: A Bussgang-Kalman Framework for Internet of Things}

\author{Chaehyun~Jung, TaeJun~Ha, Hyeonuk~Kim, and Jeonghun~Park

\thanks{This work was supported by Institute of Information $\&$ communications Technology Planning $\&$ Evaluation (IITP) under 6G Cloud Research and Education Open Hub(IITP-2025-RS-2024-00428780) grant funded by the Korea government(MSIT),
part of this work has been supported by the 6GARROW project which has received funding from the Smart Networks and Services Joint Undertaking (SNS JU) under the European Union’s Horizon Europe research and innovation programme under Grant Agreement No 101192194 and from the Institute for Information $\&$ Communications Technology Promotion (IITP) grant funded by the Korean government (MSIT) (No. RS-2024-00435652).

The authors are with School of Electrical and Electronic Engineering, Yonsei University, South Korea (e-mail: {\texttt{jch0624@yonsei.ac.kr, tjha@yonsei.ac.kr, garksi11@yonsei.ac.kr, jhpark@yonsei.ac.kr}}).
}}

\maketitle \setcounter{page}{1} 
\begin{abstract}
    Accurate state estimation from heavily quantized measurements is a key challenge in resource-constrained Internet of Things (IoT) sensing and tracking, where battery-powered devices may employ low-resolution analog-to-digital converters (ADCs) to simplify sensor hardware and reduce the amount of data.
    Existing model-based and hybrid learning-based estimators, however, typically assume high-resolution observations and therefore degrade severely under 1-bit quantization.
    In this paper, we study nonlinear state estimation with 1-bit observations and develop a Bussgang-aided filtering framework for IoT sensing front-ends with 1-bit quantization.
    For fully known system models, we propose a Bussgang-aided Kalman Filter (BKF) that explicitly incorporates quantization distortion into recursive estimation, together with a reduced-complexity variant (reduced-BKF) for computationally efficient implementation.
    For partially known models, we further propose Bussgang-aided KalmanNet (BKNet), a model-based deep learning architecture that combines adaptive thresholding with gated recurrent units (GRUs) to mitigate severe quantization effects and model mismatch.
    Experiments on the Lorenz attractor and the Michigan NCLT dataset, both under 1-bit front-end quantization, demonstrate accurate and robust state estimation under highly nonlinear dynamics, imperfect models, and extreme quantization. 
    These results support the potential of the proposed framework for reliable state estimation in resource-constrained IoT sensing and tracking applications with low-resolution front-ends.
\end{abstract} 

\begin{IEEEkeywords}
Internet of Things (IoT) sensing, state estimation, 1-bit quantization, Bussgang theorem, Kalman filtering
\end{IEEEkeywords}

%-------------------------------------------------------------------
\section{Introduction}

State estimation---inferring a hidden physical state from a sequence of noisy measurements---lies at the heart of tracking in Internet-of-Things (IoT) sensing systems. Whether a battery-powered sensor node tracks the position of a mobile robot, monitors the vibration state of an industrial asset, or follows the trajectory of a connected vehicle \cite{Dong2023GNNIoT}, the underlying task reduces to recursively combining model-based predictions with new observations so as to minimize estimation error. 
% recursively estimating a hidden state from noisy observations. 
In practice, however, IoT sensing devices must sustain years-long operation on limited battery budgets, making energy efficiency a first-order design constraint. Within the sensing front-end, the analog-to-digital converter (ADC) has been consistently identified as a dominant contributor to the overall power consumption \cite{Murmann2008ADCtrends}, while its energy cost grows exponentially with the number of quantization bits \cite{choi2022twc}. 
This motivates the adoption of 1-bit ADCs---equivalently, simple comparator-based architectures---which
%offer a radical reduction in per-measurement energy cost and hardware complexity \cite{Mollen2017LowResADC}. 
can reduce per-measurement conversion cost and hardware complexity in suitable low-resolution sensing regimes \cite{Mollen2017LowResADC}.
The resulting binary observations, however, discard all amplitude information and retain only the sign relative to a threshold, fundamentally complicating the state estimation problem that the tracking system must solve.

% it is instructive to review the state of the art in recursive state estimation, which has been developed almost exclusively under the assumption of full-resolution—or even ideal—observations.

% State estimation has been extensively studied under the assumption of ideal, full-resolution observations. 
% The classical Kalman filter (KF) \cite{kalman1960, simon2006optimal} is a representative approach for state estimation under linear-Gaussian models. By alternating between prediction and update steps, the KF recursively combines model-based prior information new observations to minimize the mean squared error (MSE).
% For nonlinear systems, extension techniques such as the extended Kalman filter (EKF) \cite{schmidt1981kalman, durbin2001series} and the unscented Kalman filter (UKF) \cite{ukf2000} have been widely adopted.
% These methods preserve the recursive structure of the KF by approximating nonlinear dynamics and uncertainties in a tractable form.
% However, their performance still depends heavily on accurate system models and observation statistics, assumptions that are often difficult to satisfy in practical IoT environments.

Under the assumption of ideal, full-resolution observations, state estimation has been extensively studied in the literature. The classical Kalman filter (KF) \cite{kalman1960, simon2006optimal} is the optimal recursive estimator under linear-Gaussian models, alternating between prediction and update steps to combine model-based prior information with new observations, so as to minimize the mean squared error (MSE). For nonlinear systems, the extended Kalman filter (EKF) \cite{schmidt1981kalman, durbin2001series} and the unscented Kalman filter (UKF) \cite{ukf2000} have been widely adopted, preserving the recursive structure of the KF by approximating nonlinear dynamics in a tractable form. 
However, their performance depends heavily on accurate system models and observation statistics, assumptions that are often difficult to satisfy in practical IoT environments. Critically, none of these methods account for the severe nonlinearity introduced by coarse quantization at the sensing front-end.

To address the dependence on fully known system models, data-driven and hybrid learning-based estimators have been actively developed recently. In \cite{rkn_icml}, the recurrent Kalman network (RKN) was proposed, wherein a DNN directly mimics the prediction and update steps of EKF. Subsequently, KalmanNet \cite{kalmannet:tsp:22} introduced a hybrid approach that preserves the recursive KF structure but learns only the Kalman gain through a DNN, rather than replacing the entire filtering process. Building upon this, Split-KalmanNet \cite{split_kal:tvt:23} improved training stability by separately learning the error covariances for the state and observation spaces. Various further extensions have since emerged: Latent KalmanNet \cite{buchnik24tsp_latent} incorporated a learning-based encoder for high-dimensional observations, Cholesky-KalmanNet \cite{ko25lsp_chol} enforced the positive-definite structure of covariance matrices, Kalmanformer \cite{shen25kalmanformer} introduced Transformer-based gain learning, and EGBRNN \cite{yan24tsp_egbrnn} addressed non-Markovian dynamics through a DNN-based memory mechanism. 
While these methods successfully relax the requirement of perfect model knowledge, they all assume access to continuous-valued, high-resolution observations. When 1-bit quantization is imposed on these imperfect models, the severe nonlinearity of the observation process introduces an additional challenge that compounds the existing difficulty of model mismatch. 
Addressing both model mismatch and quantization-induced nonlinearity within a unified recursive estimation framework remains an open problem.

A complementary line of work, which comes from low-resolution communications, particularly in 1-bit massive MIMO systems \cite{demir2021bussgang_mag, li2017bussgang, wan20twc_genBussgang, ding25tsp_mmse, yun2025uplink}, offers a promising tool: Bussgang-based linearization has been successfully used to handle 1-bit quantized measurements for channel estimation and symbol detection. 
% These works demonstrate that a statistically consistent linear surrogate can be constructed even under extreme quantization. 
% {\textcolor{blue}{
These works demonstrate that a useful second-order linear surrogate can be constructed even under extreme quantization.
% }}%
However, these results address static or block-wise parameter identification, whereas recursive state estimation requires propagating predictions and corrections over time. Extending the Bussgang framework to this dynamic regime is therefore far from straightforward.

In this paper, we address nonlinear state estimation under 1-bit quantized observations, focusing on two practical scenarios: i) the system dynamics and observation models are fully known, and ii) the system model is only partially known due to model mismatch or uncertain noise statistics. For the first scenario, we propose a Bussgang-aided Kalman filter (BKF), which incorporates the quantization distortion into a recursive filtering framework via the Bussgang decomposition. 
% A key ingredient is adaptive thresholding: by dynamically shifting the quantization threshold to center the quantizer input around zero based on the prior prediction, the BKF ensures that binary observations retain maximum information about the state variables.
% {\textcolor{blue}{
A key ingredient is adaptive thresholding: by dynamically shifting the quantization threshold according to the one-step predicted measurement, the quantizer input is made approximately zero-centered in the tracking regime.
This increases the occurrence of informative sign changes near the comparator threshold and enables a simple zero-mean Bussgang model.
% }}

% {\textcolor{blue}{
We further propose a reduced-complexity variant, termed reduced-BKF (rBKF), tailored to multi-branch sensing architectures in which multiple parallel 1-bit comparators are deployed per measurement feature. 
While such front-end structures can partially compensate for information loss due to coarse quantization, they also increase the observation dimension and the cost of the covariance inversion in BKF. 
The rBKF addresses this issue by aggregating the comparator branches associated with the same physical feature before the gain update using uniform block averaging.

For the second scenario involving partial model knowledge, we propose a Bussgang-aided KalmanNet (BKNet). Following the philosophy of model-based deep learning \cite{mbdnn:book:23}, BKNet retains the recursive prediction–correction structure of rBKF and learns only the Bussgang gain through a sequence of gated recurrent units (GRUs) that capture the latent error-covariance dynamics under both quantization distortion and model mismatch. 
This design preserves the algorithmic transparency of recursive filtering while providing robustness against imperfect models and non-Gaussian uncertainties.

% The proposed methods are validated through both synthetic and real-data experiments. On the Lorenz attractor, we simulate a 1-bit sensing front-end and evaluate estimation performance across nonlinear dynamics, various noise levels, and multiple-ADC configurations. For real-world validation, we use the Michigan NCLT dataset \cite{nclt}, where we emulate 1-bit sensing by re-quantizing recorded measurements, allowing evaluation under realistic mobile-sensing trajectories and partial model knowledge. The results demonstrate that the proposed algorithms achieve accurate state estimation even under extreme quantization. In particular, BKNet exhibits strong robustness under model mismatch, while rBKF significantly reduces inference time with negligible loss in accuracy, confirming its suitability for real-time IoT applications.

% {\textcolor{blue}{
% The proposed methods are validated through both synthetic and real-data experiments.
% The Lorenz attractor provides a controlled benchmark for 1-bit sensing, multi-branch processing, model mismatch, and diagnostic analyses of the proposed Bussgang approximation.
% The Michigan NCLT dataset \cite{nclt} is used for real-world mobile-sensing validation under emulated 1-bit measurements.
% Together, these experiments demonstrate the accuracy, robustness, and limitations of front-end-aware recursive estimation under extreme quantization.
The proposed methods are evaluated on the Lorenz attractor and the Michigan NCLT dataset under 1-bit sensing.
In addition to MSE comparisons, we further provide compact diagnostics on covariance calibration, threshold robustness, and the BKF–BKNet performance gap under real-data model mismatch.
% }}

% The performance of the proposed methods is extensively validated in both numerical simulations and real-data experiments.
% First, using the Lorenz attractor model, we simulate a 1-bit sensing front-end and analyze the estimation performance under nonlinear dynamics, various noise levels, and multiple-ADC configurations.
% For real-world validation, we use the Michigan NCLT dataset \cite{nclt}.
% Since raw pre-ADC analog waveforms are unavailable in this dataset, we emulate 1-bit sensing by re-quantizing the recorded measurements.
% This allows us to evaluate robustness under realistic mobile-sensing trajectories and partial model knowledge.
% The experimental results demonstrate that the proposed algorithms achieve accurate and robust state estimation performance even under extreme quantization.
% Notably, BKNet demonstrates highly effective under model mismatch conditions, while rBKF significantly reduces inference time with negligible degradation in accuracy, demonstrating its applicability to real-time IoT systems.

We summarize our contributions as follows.
\begin{itemize}
\item \textbf{State estimation with 1-bit sensing}: We formulate nonlinear state estimation under 1-bit quantized observations for resource-constrained IoT sensing. Unlike conventional settings, the dominant nonlinearity arises from the observation acquisition process rather than the system dynamics.
\item \textbf{BKF and rBKF}: For fully known models, we propose BKF, which integrates Bussgang decomposition with adaptive thresholding for recursive estimation under 1-bit quantization. We also present rBKF, a reduced-complexity variant that projects high-dimensional binary observations into a compact space for multi-branch 1-bit comparator front-ends.
\item \textbf{BKNet}: For partially known models, we propose BKNet, a model-based deep learning estimator that learns the Bussgang gain via sequentially connected GRUs within the rBKF structure, jointly handling quantization distortion and model mismatch.
% \item \textbf{Experimental validation}: Experiments on the Lorenz attractor and the Michigan NCLT dataset demonstrate that accurate state estimation is achievable under extreme quantization, with rBKF offering significant computational savings and BKNet exhibiting robustness under model mismatch.
% {\textcolor{blue}{
% \item \textbf{Front-end energy and architecture discussion:}
% We provide a first-order energy and latency model for the proposed adaptive-threshold comparator front-end.
% The discussion explicitly accounts for the number of comparator branches, adaptive-reference overhead, and the distinction between multi-branch 1-bit sensing and conventional multi-bit ADC architectures.
% \item \textbf{Validation, diagnostics, and real-data analysis:}
% We validate the proposed framework on the Lorenz attractor and the Michigan NCLT dataset. Beyond MSE comparisons, we provide covariance-consistency diagnostics, Gaussian-surrogate and zero-centering checks, threshold-mismatch and recovery experiments, quantization-aware SOI-KF and exact-likelihood PF baselines, and a covariance-sensitivity analysis explaining the BKF--BKNet gap on NCLT.
\item \textbf{Validation and practical considerations:} 
We compare the proposed methods against existing baselines and further provide compact analyses of covariance calibration, adaptive-threshold robustness, and front-end implementation tradeoffs.
% }}
\end{itemize}

% We summarize our contributions as follows.
% \begin{itemize}
%     \item {\textbf{State estimation with 1-bit IoT sensing front-ends}}: 
%     We formulate a nonlinear state estimation problem under 1-bit quantized observations, motivated by low-resolution sensing for resource-constrained IoT devices.
%     This problem is fundamentally different from conventional state estimation with ideal observations, as the dominant nonlinearity arises not from the system dynamics but from the observation acquisition process itself.

%     \item {\textbf{BKF and rBKF}}:
%     For fully known system models, we propose the BKF, which integrates Bussgang decomposition and adaptive thresholding into recursive state estimation.
%     We also present the rBKF, which reduces computational complexity through low-dimensional projection for multi-ADC front-ends.

%     \item {\textbf{BKNet}}:
%     For partially known system models, we propose BKNet, a model-based deep learning estimator built upon rBKF.
%     BKNet learns the Bussgang gain through a GRU-based module, enabling it to adapt to model mismatches and non-Gaussian uncertainties while maintaining a recursive filtering structure.

%     \item {\textbf{Simulation and emulation-based validation}}:
%     Through experiments on the Lorenz attractor with simulated 1-bit front-end quantization and on the Michigan NCLT dataset with emulated 1-bit sensing, we demonstrate that accurate state estimation is feasible even in heavily quantized observations, while also confirming the computational efficiency of rBKF.
% \end{itemize}

\textbf{Notations}: 
Boldface uppercase and lowercase letters denote matrices and vectors (e.g., $\mathbf{A}$, $\mathbf{a}$), respectively. 
$\mathbf{A}[i,j]$ denotes the $(i,j)$-th entry of $\mathbf{A}$.
$\mathbf{I}_N$ denotes the $N \times N$ identity matrix.
$\mathbf{1}_{N}$ denotes the $N \times 1$ vector whose elements are all equal to one.
Superscripts $(\cdot)^{\top}$ and $(\cdot)^{-1}$ denote transpose and inversion.
The operator $\operatorname{diag}(\cdot)$ creates a diagonal matrix from a vector or extracts diagonal entries from a matrix.
% {\textcolor{blue}{
$\odot$ denotes element-wise multiplication.
% }}
$\mathbb{R}$ and $\mathbb{N}$ denote the real numbers and natural numbers, respectively.
The Gaussian distribution with mean $\mu$ and covariance $\mathbf{\Sigma}$ is denoted as $\mathcal{N}(\mu, \mathbf{\Sigma})$. 
The subscript notation $(\cdot)_{t}$ denotes data at time $t$, while $(\cdot)_{t|t-1}$ and $(\cdot)_{t|t}$ denote, respectively, the prediction and the estimate at time $t$ based on observations up to time $t-1$ and $t$.

%-------------------------------------------------------------------
\section{System Model and Problem Formulation}
%------------------------------------------------------------------

\begin{figure*}[t]
    \centering
    \includegraphics[width=0.9\textwidth]{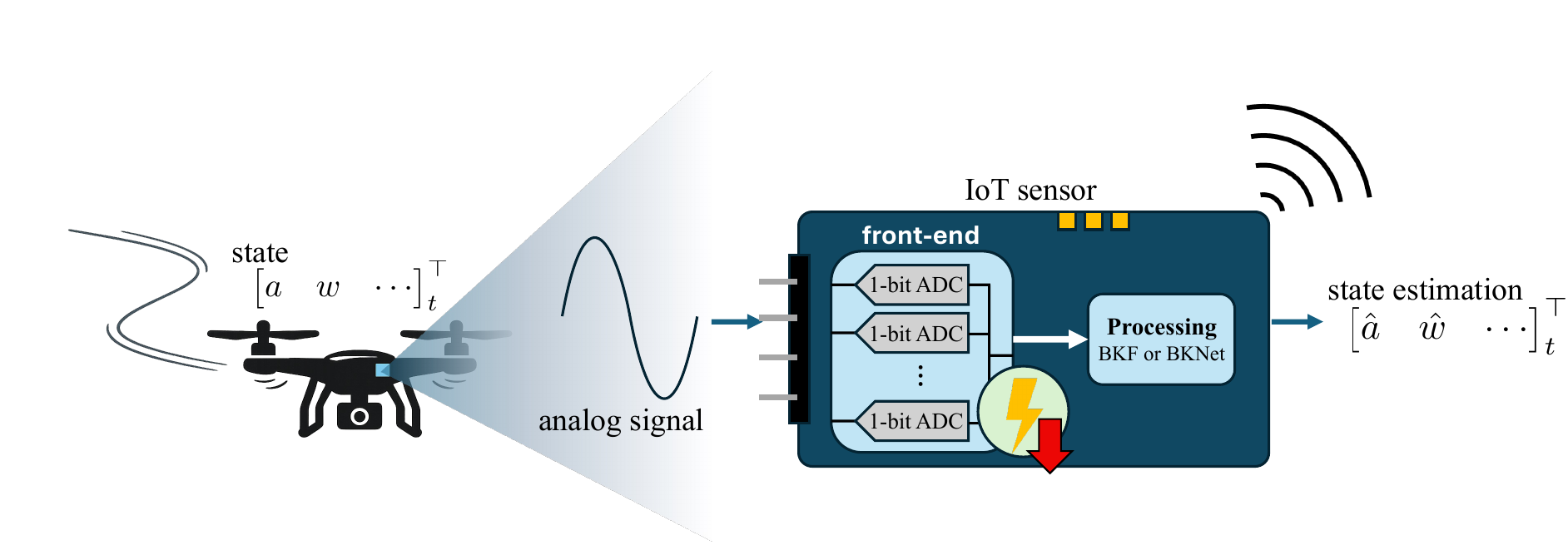}
    \caption{Illustration of the considered IoT sensing architecture. A resource-constrained IoT sensor acquires analog signals related to the latent state and digitizes them through a 1-bit front-end, motivated by hardware simplicity and energy-efficient operation. The resulting binary measurements are then processed by BKF or BKNet to estimate the state.}
    \label{fig:iot_architecture}
\end{figure*}

In this section, we describe the IoT sensing architecture, the nonlinear SS model, and estimation problem considered in this paper.
We consider a resource-constrained IoT sensor node that observes a latent physical state, forms a pre-quantization measurement vector, digitizes it through a sensing front-end, and forwards the resulting digital output to a state estimator.
Depending on the sensing-front-end resolution, we distinguish two scenarios: i) a conventional high-resolution ADC, which is used as a reference model, and ii) a 1-bit comparator-based front-end, which is the primary focus of this work.
The latter captures low-resolution sensing architectures motivated by hardware simplicity and reduced measurement-data volume.
This overall sensing-and-estimation pipeline is illustrated in Fig.~\ref{fig:iot_architecture}.

%-------------------------------------------------------------------
\subsection{Baseline State Estimation with High-Resolution Sensing Front-end}

We first consider the discrete-time nonlinear SS model with a high-resolution sensing front-end.
The latent state evolves as
\begin{align}
    \mathbf{x}_t = \mathbf{f}(\mathbf{x}_{t-1}) + \mathbf{w}_t,
\end{align}
and the corresponding measurement is given by:
\begin{align}
    \mathbf{y}_t = \mathbf{h}(\mathbf{x}_t) + \mathbf{v}_t,
\end{align}
where $\mathbf{x}_t \in \mathbb{R}^m$ is the state vector, $\mathbf{f}: \mathbb{R}^m \rightarrow \mathbb{R}^m$ is the state transition function, $\mathbf{h}: \mathbb{R}^m \rightarrow \mathbb{R}^n$ is the measurement function, and $\mathbf{w}_t \in \mathbb{R}^m$ and $\mathbf{v}_t \in \mathbb{R}^n$ denote the process and measurement noise, respectively.
For modeling convenience, we typically assume $\mathbf{w}_t \sim \mathcal{N}(\mathbf{0}, \mathbf{Q}_t)$ and $\mathbf{v}_t \sim \mathcal{N}(\mathbf{0}, \mathbf{R}_t)$.
Furthermore, the SS model satisfies the Markov property, as $\mathbf{y}_t$ depends only on the current state $\mathbf{x}_t$, which in turn depends only on the previous state $\mathbf{x}_{t-1}$.
The objective of the state estimation problem is to solve the following optimization:
\begin{align} \label{eq:prob_formulation_hr}
    \underset{\hat{\mathbf{x}}_t}{\text{minimize}} \;\; \mathbb{E} \! \left[ \|\mathbf{x}_t - \hat{\mathbf{x}}_t \|^2 \mid \mathbf{y}_{1:t} \right].
\end{align}

When $\mathbf{f}$, $\mathbf{h}$, and the relevant statistics are known, the EKF provides a standard recursive approximation for \eqref{eq:prob_formulation_hr}. 
Specifically, in the prediction step, it computes:
\begin{align}
    &\hat{\mathbf{x}}_{t|t-1} = \mathbf{f}(\hat{\mathbf{x}}_{t-1|t-1}), \\
    &\mathbf{\Sigma}_{t|t-1} = \mathbf{F}_t \mathbf{\Sigma}_{t-1|t-1} \mathbf{F}_t^{\top} + \mathbf{Q}_t, \\
    &\hat{\mathbf{y}}_{t|t-1} = \mathbf{h}(\hat{\mathbf{x}}_{t|t-1}), \\
    &\mathbf{P}_{t|t-1} = \mathbf{H}_t \mathbf{\Sigma}_{t|t-1} \mathbf{H}_t^{\top} + \mathbf{R}_t,
\end{align}
where $\mathbf{F}_t$ and $\mathbf{H}_t$ denote the Jacobians of $\mathbf{f}(\cdot)$ and $\mathbf{h}(\cdot)$, respectively.
Subsequently, the update step is expressed as:
\begin{align}
    &\mathbf{KG}_{t} = \mathbf{\Sigma}_{t|t-1}\mathbf{H}^{\top}_t \mathbf{P}^{-1}_{t|t-1}, \label{eq:EKF_KG} \\
    &\hat{\mathbf{x}}_{t|t} = \hat{\mathbf{x}}_{t|t-1} + \mathbf{KG}_t \big(\mathbf{y}_t - \hat{\mathbf{y}}_{t|t-1}\big), \\
    &\mathbf{\Sigma}_{t|t} = \mathbf{\Sigma}_{t|t-1} - \mathbf{KG}_t \mathbf{P}_{t|t-1}\mathbf{KG}_{t}^{\top}.
\end{align}
If $\mathbf{f}$ and $\mathbf{h}$ are linear, the EKF reduces to the conventional KF, which is optimal for linear-Gaussian models in the MSE sense.

Meanwhile, in scenarios where the system model or noise statistics are only partially known, hybrid estimators such as KalmanNet \cite{kalmannet:tsp:22} apply a learned gain instead of an analytical Kalman gain, with the update rule given by:
\begin{align}
    \hat{\mathbf{x}}_{t|t} = \hat{\mathbf{x}}_{t|t-1} + \CMcal{KG}_t(\Theta) \big( \mathbf{y}_t - \hat{\mathbf{y}}_{t|t-1} \big),
\end{align}
where $\CMcal{KG}_t (\Theta)$ denotes the learned Kalman gain parameterized by trainable parameters $\Theta$.
Related to this, Split-KalmanNet \cite{split_kal:tvt:23} can be viewed as a split-gain-learning variant of KalmanNet.
Specifically, Split-KalmanNet constructs the Kalman gain as:
\begin{align}
    \CMcal{KG}^{\text{split}}_t(\Theta_{\mathbf{\Sigma}},\Theta_{\mathbf{P}},\mathbf{H}_{t}) = \CMcal{G}_{t,\mathbf{\Sigma}}(\Theta_{\mathbf{\Sigma}}) \, \mathbf{H}_t \, \CMcal{G}_{t,\mathbf{P}}(\Theta_{\mathbf{P}}),
\end{align}
where $\CMcal{G}_{t,\mathbf{\Sigma}}(\Theta_{\mathbf{\Sigma}})$ and $\CMcal{G}_{t,\mathbf{P}}(\Theta_{\mathbf{P}})$ represent recurrent modules that separately learn the covariance-related representations of the state and observation spaces, respectively.
This split structure enables more robust estimation under model mismatches.

An important point is that EKF, KalmanNet, and Split-KalmanNet are all designed under the assumption of high-resolution measurements $\mathbf{y}_t$.
That is, the sensing front-end is assumed to introduce no explicit quantization distortion beyond measurement noise.

%-------------------------------------------------------------------
\subsection{State Estimation with 1-Bit IoT Sensing Front-ends} \label{section:1-bit_state_estimation}

We now formulate the low-resolution sensing problem that is the primary focus of this paper.
Let $\mathbf{y}_t$ denote the pre-quantization measurement vector generated by the sensing front-end.
The 1-bit digital observation delivered to the state estimator is modeled as:
\begin{align}
    \mathbf{r}_t = \mathcal{Q}\big(\mathbf{y}_t - \boldsymbol{\tau}_t\big) 
    = \mathcal{Q}(\mathbf{z}_t), \qquad \mathbf{z}_t = \mathbf{y}_t-\boldsymbol{\tau}_t.
\end{align}
where $\boldsymbol{\tau}_t \in \mathbb{R}^n$ is the front-end threshold vector.
% {\textcolor{blue}{
Here, $\mathcal{Q}: \mathbb{R}^n \to \{-1,+1\}^n$ denotes the vector-valued 1-bit quantizer obtained by applying the scalar comparator $q(\cdot)$ element-wise.
We also use bracket notation such as $\mathbf{z}[i]$ to denote vector entries.
% \begin{align}
%     \mathbf{Q}(\mathbf{z})[i] =
%     \begin{cases}
%          1,  & \mathbf{z}[i] > 0, \\
%          -1, & \mathbf{z}[i] \le 0.
%     \end{cases}
% \end{align}
\begin{equation}\begin{aligned}
    &\mathcal{Q}(\mathbf{z}) \triangleq \big[ q(\mathbf{z}[1]),q(\mathbf{z}[2]), \ldots, q(\mathbf{z}[n]) \big]^\top, \\
    &q(u) = \begin{cases}
    1,  & u>0, \\
    -1, & u\le 0.
    \end{cases}
\end{aligned}\end{equation}
The zero-threshold case is obtained by setting $\boldsymbol{\tau}_t = 0$.
More generally, $\boldsymbol{\tau}_t$ can be fixed or time varying.
In the adaptive threshold setting considered later, $\boldsymbol{\tau}_t$ is determined from the one-step predicted measurement, and is therefore known to both the front-end and the estimator at time $t$.
% This reference update does not require access to the current high-resolution measurement after quantization; it only adjusts the comparator threshold before the 1-bit decision is made.
% }}

% {\textcolor{blue}{
We now clarify the single-branch and multi-branch dimensions.
To avoid ambiguity in the multi-branch setting, let $n_0$ be the number of physical measurement features and $L$ the number of 1-bit comparator branches per feature.
% The aggregate pre-quantization vector is
% \begin{align}
%     \mathbf{y}_t = \tilde{\mathbf{h}}(\mathbf{x}_t) \otimes \mathbf {1}_L + \mathbf{v}_t \in \mathbb{R}^{Ln_0},
% \end{align}
% so the binary observation dimension is $n=Ln_0$.
% The output of the $\ell$-th branch for feature $j$ is
% \begin{align}
%     \mathbf{r}_t[j,\ell] = q(\mathbf{y}_t[j,\ell] - \boldsymbol{\tau}_t[j,\ell]).
% \end{align}
% For adaptive thresholding, we use a shared feature-level reference,
% \begin{align}
%     \boldsymbol{\tau}_t = \bar{\boldsymbol{\tau}}_t \otimes \mathbf{1}_L,
% \end{align}
% so the adaptive component is shared across branches observing the same physical feature.
% Let $n_0$ denote the number of physical measurement features, and let
% \begin{align}
%     \tilde{\mathbf{y}}_t = \tilde{\mathbf{h}} (\mathbf{x}_t) + \tilde{\mathbf{v}}_t \in \mathbb{R}^{n_0}
% \end{align}
% be the underlying pre-quantization measurement feature vector.
In the single-branch case, $L=1$, $n=n_0$, and $y_t=\tilde{y}_t$.
In the multi-branch case, $L$ parallel 1-bit comparator branches are used for each physical feature, so the aggregate pre-quantization vector is
\begin{align}
    \mathbf{y}_t = \tilde{\mathbf{h}}(\mathbf{x}_t)\otimes \mathbf{1}_L + \mathbf{v}_t \in \mathbb{R}^{n},
    \qquad
    n=L n_0.
\end{align}
Equivalently, the scalar output of the $\ell$-th branch associated with the $j$-th physical feature is
\begin{equation}\begin{aligned}
    &\mathbf{r}_t[j,\ell] = q\!\left( \mathbf{y}_t[j,\ell] - \boldsymbol{\tau}_t[j,\ell] \right), \\
    &j = 1, \ldots, n_0,
    \quad
    \ell = 1, \ldots, L.
\end{aligned}\end{equation}
% The threshold can be written as
% \begin{align}
%     \boldsymbol{\tau}_t = \bar{\boldsymbol{\tau}}_t \otimes \mathbf{1}_L,
% \end{align}
% where $\bar{\boldsymbol{\tau}}_t \in \mathbb{R}^{n_0}$ is the feature-level adaptive threshold.
% Thus, the adaptive component of the threshold is shared across comparator branches observing the same physical feature.
% }}%

% The zero-threshold case is obtained by setting $\pmb{\tau}_t = 0$.
% More generally, $\pmb{\tau}_t$ can be fixed or time varying, and the latter will later enable adaptive thresholding or dithering strategies.
% For clarity, we refer to $\mathbf{y}_t$ as the pre-quantization measurement and $\mathbf{r}_t$ as the 1-bit observation.
% In the single-ADC setting, the observation dimension $n$ corresponds to the number of measurement features.
% However, in the multi-ADC setting explored subsequently, a single underlying measurement feature may be digitized through multiple parallel comparator branches.
% In this case, $n$ represents the aggregate dimension of the resulting binary outputs, meaning that multiple elements of $\mathbf{r}_t$ may correspond to the same underlying measurement feature.

Under this front-end model, the state estimation problem is reformulated as:
\begin{align} \label{eq:prob_formulation_1bit}
    \underset{\hat{\mathbf{x}}_{t|t}}{\text{minimize}} \;\; \mathbb{E} \! \left[ \|\mathbf{x}_t - \hat{\mathbf{x}}_{t|t}\|^2 \mid \mathbf{r}_{1:t} \right].
\end{align}
Compared with the high-resolution case, the estimator now receives only the sign of each thresholded measurement.
As a result, amplitude information is lost, and the observation statistics become highly non-Gaussian, discontinuous, and threshold dependent.
Thus, the standard EKF and KalmanNet recursions are not directly compatible with this sensing model.
More specifically, their update rules rely on innovations formed from high-resolution measurement residuals or smoothly modeled observation mappings, failing to explicitly capture the nonlinear distortion induced by the 1-bit front-end.
This necessitates a recursive estimator that is explicitly quantization-aware, which serves as the core motivation for the BKF, rBKF, and BKNet proposed in this paper.

% {\color{blue}{
\begin{remark}[Front-end interpretation] \normalfont

The adaptive threshold $\boldsymbol{\tau}_t$ is a programmable comparator reference set from the one-step predicted measurement computed using past 1-bit observations before the current decision; it does not require access to the current unquantized measurement.
Accordingly, the proposed front-end is an algorithmic abstraction rather than a measured circuit implementation.
A conventional $b$-bit ADC produces a multi-level amplitude codeword, whereas the shared-reference $L$-branch front-end considered here produces $L$ binary comparisons.
Its first-order per-sample energy can be modeled as
\begin{align}
    E_{\mathrm{1b}}(L) \approx L E_{\mathrm{cmp}} + E_{\tau} + E_{\mathrm{ref}} + E_{\mathrm{IO}}^{\mathrm{1b}}(L),
\end{align}
where the non-comparator terms represent threshold-update, reference-generation/settling, and binary-I/O overheads.
For comparison, the conversion-core energy of a conventional ADC follows
$E_{\mathrm{ADC},b}\approx\mathrm{FoM}_{\mathrm{W}}(b)2^b$~\cite{Murmann2008ADCtrends}.
Thus, the potential energy advantage of 1-bit sensing decreases with $L$, adaptive-reference overhead, and I/O cost, and is most favorable in the single-branch or small-$L$ regime.
The proposed front-end should therefore be viewed as an implementation-dependent accuracy--complexity--energy tradeoff, not as a universal energy-efficiency claim over multi-bit ADCs.
\end{remark}

\section{Bussgang-aided Kalman Filter}
%-------------------------------------------------------------------

In this section, we develop the Bussgang-aided Kalman filter (BKF) to solve \eqref{eq:prob_formulation_1bit} for a 1-bit comparator-based IoT sensing front-end.
Our objective is to construct a recursive estimator that explicitly accounts for front-end quantization while retaining the prediction and update structure of Kalman filtering.
Throughout this section, we assume that the state transition and measurement models, i.e., $\mathbf{f}(\cdot)$, $\mathbf{h}(\cdot)$, and the distribution of $\mathbf{w}_t$ and $\mathbf{v}_t$, are fully known.
This assumption will be relaxed later in the development of BKNet.
% {\textcolor{blue}{
We distinguish the exact second-order Bussgang relation from the Gaussian surrogate used only to close the recursive covariance update.
% ; the consistency of this surrogate is evaluated empirically in Section~\ref{section:numerical_exp}.
% }}

%-------------------------------------------------------------------
\subsection{Zero-Centered Bussgang Linearization}

The Bussgang theorem \cite{Bussgang1952} states that for a zero-mean Gaussian input and a memoryless nonlinear output, the input and output cross-correlation is proportional to the input autocorrelation.
% Specifically, consider a zero-mean Gaussian vector $\mathbf{y}$ with a covariance matrix $\mathbf{C}_{\mathbf{y}}$ and its element-wise 1-bit quantized output $\mathbf{r} = \mathcal{Q}(\mathbf{y})$.
For notational simplicity, the time index is omitted in this part.
% Then, according to the Bussgang theorem:
% \begin{align} \label{eq:buss_theorem}
%     \mathbf{C}_{\mathbf{ry}} = \mathbf{B} \mathbf{C}_{\mathbf{y}},
% \end{align}
% where $\mathbf{C}_{\mathbf{ry}} \triangleq \mathbb{E}[\mathbf{r}\mathbf{y}^{\top}]$, and the Bussgang coefficient matrix is given by
% \begin{align} \label{eq:buss_coefficient}
%     \mathbf{B} = \sqrt{\frac{2}{\pi}}\operatorname{diag}(\mathbf{C}_{\mathbf{y}})^{-\frac{1}{2}}.
% \end{align}
% Consequently, the 1-bit output can be represented by the following linearized model:
% \begin{align}
%     \mathbf{r} = \mathcal{Q}(\mathbf{y}) = \mathbf{B} \mathbf{y} + \pmb{\eta},
% \end{align}
% where $\pmb{\eta}$ denotes an effective zero-mean distortion term that is uncorrelated with $\mathbf{y}$.

% {\color{blue}{
Consider a zero-mean Gaussian vector $\mathbf{z}\in\mathbb{R}^n$ with covariance matrix $\mathbf{C}_\mathbf{z}$, and its component-wise 1-bit quantized output $\mathbf{r} = \mathcal{Q}(\mathbf{z})$.
The Bussgang relation gives
\begin{align}
    \mathbf{C}_{\mathbf{rz}} = \mathbf{B}\mathbf{C}_\mathbf{z}, \label{eq:buss_theorem}
\end{align}
where $\mathbf{C}_{\mathbf{rz}} \triangleq \mathbb{E}[\mathbf{r}\mathbf{z}^{\top}]$.
For the sign quantizer, the Bussgang matrix is diagonal and is given by
\begin{align}
    \mathbf{B} = \sqrt{\frac{2}{\pi}}\, \operatorname{diag}(\mathbf{C}_\mathbf{z})^{-1/2}, \label{eq:buss_coefficient}
\end{align}
where $\operatorname{diag}(\mathbf{C}_\mathbf{z})$ denotes the diagonal matrix formed from the diagonal entries of $\mathbf{C}_\mathbf{z}$.

Defining the effective distortion as $\boldsymbol{\eta} = \mathbf{r}-\mathbf{B}\mathbf{z}$,
we obtain
\begin{align}
    \mathbb{E}[\boldsymbol{\eta}\mathbf{z}^{\top}]
    = \mathbb{E}[(\mathbf{r} - \mathbf{B}\mathbf{z}) \mathbf{z}^{\top}]
    = \mathbf{C}_{\mathbf{rz}} - \mathbf{B}\mathbf{C}_\mathbf{z}
    = \mathbf{0}.
\end{align}
Therefore, the 1-bit output can be represented as
\begin{align}
    \mathbf{r} = \mathcal{Q}(\mathbf{z})
    = \mathbf{B}\mathbf{z} + \boldsymbol{\eta},
\end{align}
where $\boldsymbol{\eta}$ is uncorrelated with $\mathbf{z}$.
Importantly, the Bussgang relation guarantees this uncorrelatedness property, but it does not imply that $\boldsymbol{\eta}$ is independent of $\mathbf{z}$ or exactly Gaussian.
% }}

The generalized Bussgang linearization for quantizers with non-zero mean inputs is detailed in Appendix~\ref{app:nonzero_bussgang}.
As demonstrated therein, both the effective Bussgang gain and the covariance of the quantized output become dependent on the input mean, which can substantially increase the computational overhead.
This serves as the primary motivation for adopting the zero-centering threshold design in the proposed BKF.

Specifically, we define the threshold quantizer input as:
\begin{align}
    &\mathbf{z}_t = \mathbf{y}_t - \boldsymbol{\tau}_t, \\
    &\mathbf{r}_t = \mathcal{Q}(\mathbf{z}_t), 
\end{align}
and choose the threshold vector based on the one-step predicted measurement mean:
\begin{align}
    \boldsymbol{\tau}_t
    = \mathbb{E} \! \left[ \mathbf{y}_t \mid \mathbf{r}_{1:t-1} \right] 
    \approx \hat{\mathbf{y}}_{t|t-1}. 
\end{align}
By this choice, the quantizer input becomes approximately zero-centered, i.e.,
\begin{align}
    \mathbb{E} \! \left[ \mathbf{z}_t \mid \mathbf{r}_{1:t-1} \right] \approx \mathbf{0}. \label{eq:z_mean}
\end{align}

% {\color{blue}{
For a multi-branch front-end with $L$ comparators per physical measurement feature, the same feature-level predicted measurement is replicated across the corresponding branches.
\begin{align}
    \boldsymbol{\tau}_t = \hat{\tilde{\mathbf{y}}}_{t|t-1} \otimes \mathbf{1}_L.
\end{align}
Thus, the adaptive component of the threshold is shared by the branches measuring the same physical feature.
Branch-specific threshold offsets or dithers can be incorporated through the generalized non-zero-mean formulation in Appendix~\ref{app:nonzero_bussgang}, but the core BKF derivation below uses the zero-centered construction.
This zero-centering not only simplifies the Bussgang decomposition but also increases the occurrence of informative sign changes near the comparator threshold.
Since \eqref{eq:z_mean} relies on the quality of the one-step prediction, it should be interpreted as an approximation rather than an exact identity.
Its validity and robustness under threshold mismatch and transient tracking errors are evaluated in Section~\ref{section:numerical_exp}.

Applying \eqref{eq:buss_theorem} and \eqref{eq:buss_coefficient} to the approximately zero-centered input $\mathbf{z}_t$ with conditional covariance $\mathbf{P}_{t|t-1}$, we obtain the time-varying Bussgang surrogate
\begin{align}
    &\mathbf{r}_t = \mathcal{Q}(\mathbf{z}_t)
    = \mathbf{B}_t\mathbf{z}_t+\boldsymbol{\eta}_t, \label{eq:linearized_r} \\
    &\mathbf{B}_t = \sqrt{\frac{2}{\pi}}\, \operatorname{diag}(\mathbf{P}_{t|t-1})^{-1/2},
\end{align}
where $\boldsymbol{\eta}_t$ is uncorrelated with $\mathbf{z}_t$ under the zero-centered Bussgang model.
% }}
% This zero-centering not only simplifies the Bussgang decomposition but also ensures that informative sign variations occur more frequently near the comparator threshold.

% Applying \eqref{eq:buss_theorem}, \eqref{eq:buss_coefficient} to the zero-centered input $\mathbf{z}_t$ with covariance $\mathbf{P}_{t|t-1}$, we obtain:
% \begin{align}
%     &\mathbf{r}_t = \mathcal{Q}(\mathbf{z}_t) = \mathbf{B}_t \mathbf{z}_t + \pmb{\eta}_t, \label{eq:linearized_r} \\
%     &\mathbf{B}_t = \sqrt{\frac{2}{\pi}}\operatorname{diag}(\mathbf{P}_{t|t-1})^{-\frac{1}{2}},
% \end{align}
% where $\pmb{\eta}_t$ is uncorrelated with $\mathbf{z}_t$.
% This linearized observation model is the key building block of the proposed BKF.

%-------------------------------------------------------------------
\subsection{Bussgang-aided Kalman Filter}

\begin{figure*}[t]
    \centering
    \subfigure[Overall filtering architecture]{
        \includegraphics[width=0.7\textwidth]{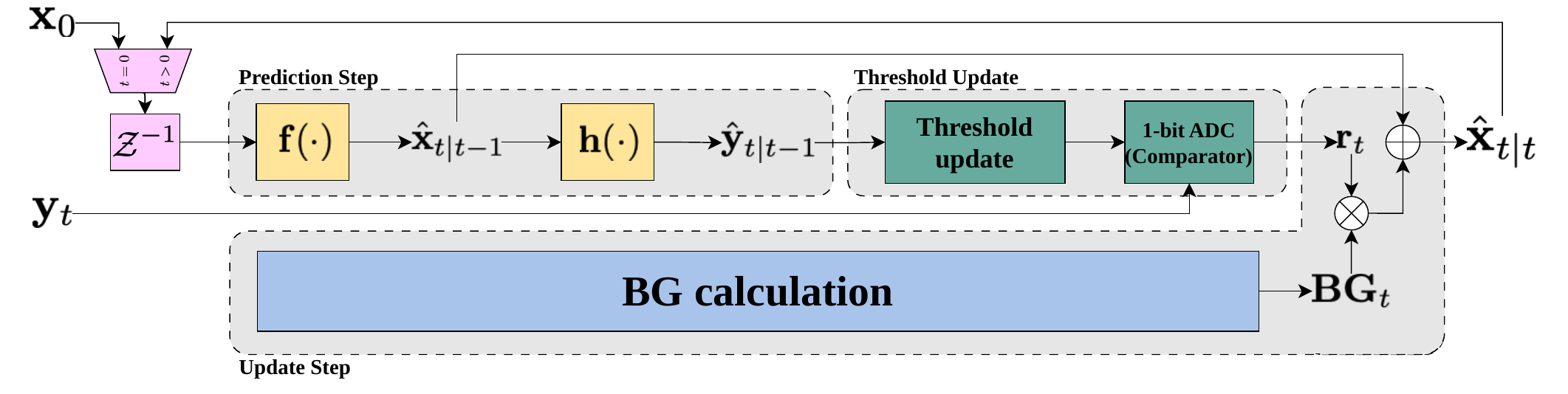}
        \label{fig:BussgangKalman}}
    \\[1ex]
    \subfigure[Bussgang gain calculation for BKF]{
        \includegraphics[width=0.7\textwidth]{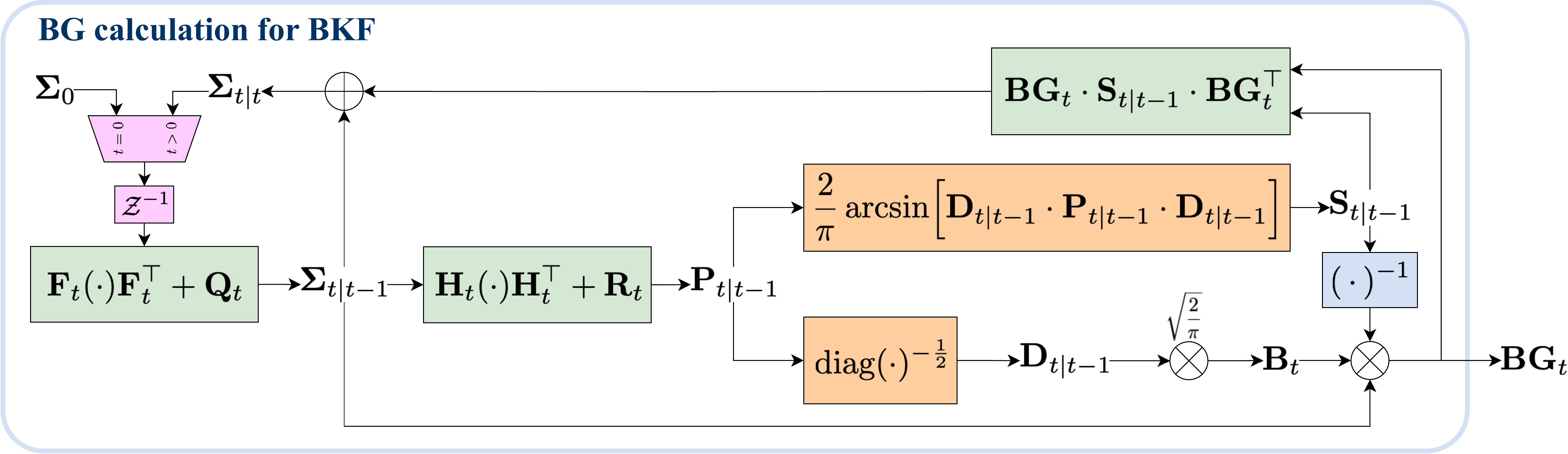}
        \label{fig:BKF}}
    \\[1ex]
    \subfigure[Bussgang gain calculation for BKNet]{
        \includegraphics[width=0.7\textwidth]{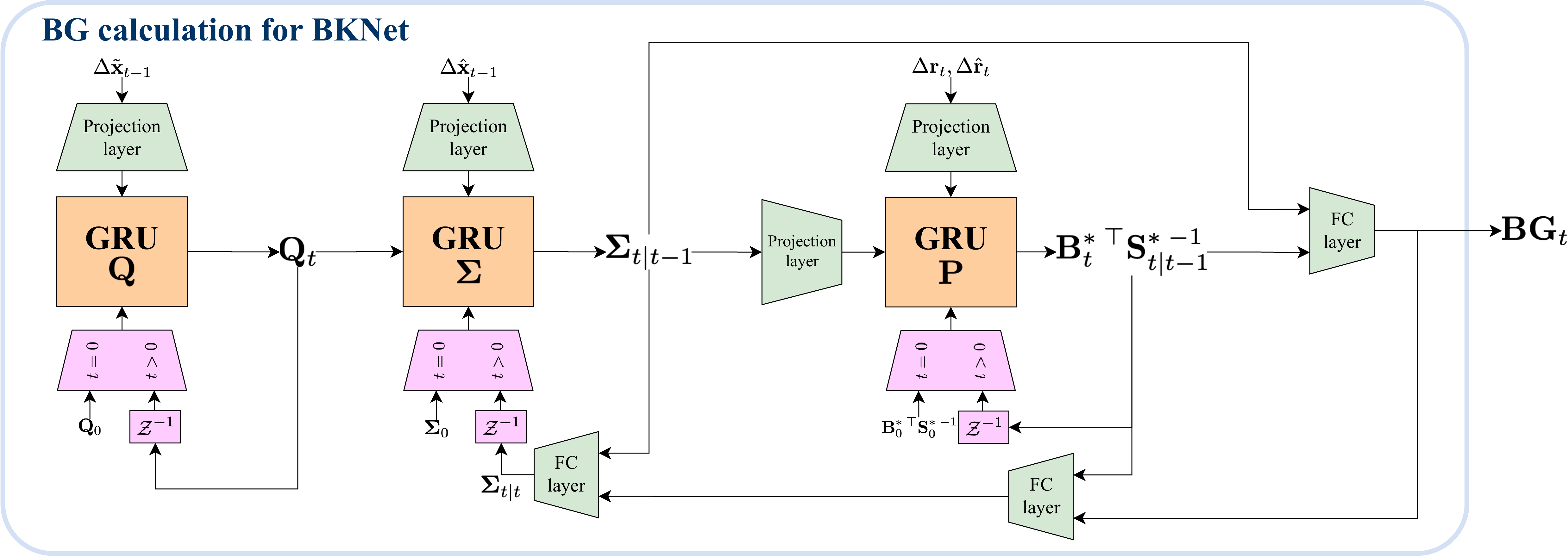}
        \label{fig:BKN}}
    \caption{(a) Overall architecture with (b) BKF and (c) BKNet.}
    \label{fig:BussgangKalman_architecture}
\end{figure*}

% Based on the zero-centered construction above, the BKF first forms the quantizer input as $\mathbf{z}_t = \mathbf{y}_t - \hat{\mathbf{y}}_{t|t-1}$ and then applies the Bussgang linearization in \eqref{eq:linearized_r}.
% For analytical tractability, we further approximate the effective distortion $\pmb{\eta}_t$ as zero-mean Gaussian noise with second-order statistics matched to the linearized model.
% {\textcolor{blue}{
Based on the zero-centered construction above, the front-end forms the thresholded quantizer input $\mathbf{z}_t = \mathbf{y}_t - \boldsymbol{\tau}_t$, while the estimator receives only $\mathbf{r}_t = \mathcal{Q}(\mathbf{z}_t)$.
BKF uses the corresponding Bussgang surrogate in \eqref{eq:linearized_r}.
The Bussgang theorem provides the second-order cross-correlation structure of the quantized observation, but it does not specify the full posterior distribution.
Therefore, BKF uses a Gaussian surrogate with matched second-order statistics to close the recursive filtering equations.
The resulting covariance should be interpreted as the covariance propagated by this Bussgang-Gaussian surrogate.
% , whose empirical consistency is examined in Section~\ref{section:numerical_exp}.
% }}%
As a result, the BKF is composed of the following prediction and update steps.

\begin{enumerate}
    \item \textbf{Prediction step}:
    Given the posterior state moments at time $t-1$, the BKF computes the prior state moments as:
    \begin{align}
        &\hat{\mathbf{x}}_{t|t-1} = \mathbf{f}(\hat{\mathbf{x}}_{t-1|t-1}), \\
        &\mathbf{\Sigma}_{t|t-1} = \mathbf{F}_t \mathbf{\Sigma}_{t-1|t-1} \mathbf{F}^{\top}_t + \mathbf{Q}_t, 
    \end{align}
    where $\mathbf{F}_t$ is the Jacobian of $\mathbf{f}(\cdot)$ evaluated at $\hat{\mathbf{x}}_{t-1|t-1}$.
    The corresponding predicted measurement moments are given by:
    \begin{align}
        &\hat{\mathbf{y}}_{t|t-1} = \mathbf{h}(\hat{\mathbf{x}}_{t|t-1}), \\
        &\mathbf{P}_{t|t-1} = \mathbf{H}_t \mathbf{\Sigma}_{t|t-1} \mathbf{H}^{\top}_t + \mathbf{R}_t,
    \end{align}
    where $\mathbf{H}_t$ is the Jacobian of $\mathbf{h}(\cdot)$ evaluated at $\hat{\mathbf{x}}_{t|t-1}$.
    Since $\mathbf{z}_t = \mathbf{y}_t - \hat{\mathbf{y}}_{t|t-1}$, its prior moments satisfy:
    % \begin{align}
    %    &\hat{\mathbf{z}}_{t|t-1} = \mathbf{0}, \quad
    %    &\mathbf{Z}_{t|t-1} = \mathbf{P}_{t|t-1}.
    % \end{align}
    % {\color{blue}{
    \begin{align}
        &\hat{\mathbf{z}}_{t|t-1} =\mathbb{E}[\mathbf{z}_t\mid \mathbf{r}_{1:t-1}] \approx \mathbf{0}, \\
        &\operatorname{Cov}(\mathbf{z}_t\mid \mathbf{r}_{1:t-1}) = \mathbf{P}_{t|t-1}.
    \end{align}
    Since $\boldsymbol{\tau}_t$ is determined from past observations, it is fixed under the conditioning on $\mathbf{r}_{1:t-1}$.
    Thus, the shift $\mathbf{z}_t = \mathbf{y}_t - \boldsymbol{\tau}_t$ changes the conditional mean of $\mathbf{y}_t$, but not its conditional covariance.
    Hence, $\mathbf{P}_{t|t-1}$ is the predicted covariance of the pre-quantization measurement, while the approximate zero-mean condition is used to obtain the simplified Bussgang gain and arcsine-law covariance of the 1-bit output.
    
    Define $\mathbf{D}_{t|t-1} = \operatorname{diag}(\mathbf{P}_{t|t-1})^{-1/2}$ and the corresponding correlation matrix
    \begin{align}
    \mathbf{C}_{t|t-1} = \mathbf{D}_{t|t-1} \mathbf{P}_{t|t-1} \mathbf{D}_{t|t-1}.
    \end{align}
    Under the zero-centered Gaussian surrogate, the prior mean of the 1-bit observation and the prior covariance of the 1-bit observation are
    \begin{align}
    &\hat{\mathbf{r}}_{t|t-1} = \mathbb{E}[\mathcal{Q}(\mathbf{z}_t) \mid \mathbf{r}_{1:t-1}] \approx \mathbf{0}, \label{eq:r_mean} \\
    &\mathbf{S}_{t|t-1} = \frac{2}{\pi} \arcsin(\mathbf{C}_{t|t-1}).
    \end{align}
    Here, $\arcsin(\cdot)$ is applied element-wise to the entries of the correlation matrix $\mathbf{C}_{t|t-1}$.
    % }}%
    This covariance follows from the arcsine law for zero-mean Gaussian sign quantization and can equivalently be obtained via Price's theorem \cite{prices_theorem}.
    % {\textcolor{blue}{
    The conditioning in \eqref{eq:r_mean} indicates that the expectation is taken with respect to the prior distribution at time $t$, after the threshold has been determined from past observations.
    % }}
    % Defining $\mathbf{D}_{t|t-1} = \operatorname{diag}(\mathbf{P}_{t|t-1})^{-\frac{1}{2}}$, the corresponding prior moments of the 1-bit observation are obtained as:
    % \begin{align}
    %     &\hat{\mathbf{r}}_{t|t-1} = \mathbb{E}[\mathcal{Q}(\mathbf{z}_t)] = \mathbf{0}, \\
    %     &\mathbf{S}_{t|t-1} = \frac{2}{\pi}\arcsin \! \left(\mathbf{D}_{t|t-1} \mathbf{P}_{t|t-1} \mathbf{D}_{t|t-1}\right),
    % \end{align}
    % where $\arcsin(\cdot)$ is applied element-wise.
    % This covariance is derived from the arcsine-law for zero-mean Gaussian sign quantization and can equivalently be obtained via Price's theorem \cite{prices_theorem}.
    
    \item \textbf{Update step}:
    Using the predicted moments and linearized observation model, the BKF updates the state moments as:
    \begin{align}
        &\hat{\mathbf{x}}_{t|t} = \hat{\mathbf{x}}_{t|t-1} + \mathbf{BG}_t \big( \mathbf{r}_t - \hat{\mathbf{r}}_{t|t-1} \big), \label{eq:BKF_xhat} \\
        &\mathbf{\Sigma}_{t|t} = \mathbf{\Sigma}_{t|t-1} - \mathbf{BG}_t \mathbf{S}_{t|t-1} \mathbf{BG}^{\top}_t, 
    \end{align}
    where the Bussgang gain is given by:
    \begin{align}
        \mathbf{BG}_t = \mathbf{\Sigma}_{t|t-1} (\mathbf{B}_t \mathbf{H}_t)^{\top} \mathbf{S}^{-1}_{t|t-1}. \label{eq:BKF_BGain}
    \end{align}
    Equivalently, under the local Gaussian prior approximation
    \begin{align}
        \mathbf{z}_t \approx \mathbf{H}_t \mathbf{e}_{t|t-1} + \mathbf{v}_t
        = \mathbf{H}_t(\mathbf{x}_t - \hat{\mathbf{x}}_{t|t-1}) + \mathbf{v}_t,
    \end{align}
    the Bussgang relation gives
    \begin{align}
        \mathrm{Cov}(\mathbf{e}_{t|t-1}, \mathbf{r}_t) = \mathbf{\Sigma}_{t|t-1} (\mathbf{B}_t \mathbf{H}_t)^\top.
    \end{align}
    Therefore, \eqref{eq:BKF_BGain} is the moment-matched LMMSE correction gain based on the first- and second-order moments of the 1-bit observation.
    This interpretation does not require the effective distortion itself to be Gaussian; Gaussianity is used only for tractable recursive moment propagation.
    % }}%
    Since $\hat{\mathbf{r}}_{t|t-1}=\mathbf{0}$ in the zero-centered construction, the correction term reduces to $\mathbf{BG}_t \mathbf{r}_t$.
\end{enumerate}

Conceptually, the BKF performs a role analogous to the EKF, distinguished primarily by the nature of the dominant nonlinearity it addresses.
While the EKF linearizes the nonlinear SS model through Jacobian-based approximations, the BKF linearizes the 1-bit sensing front-end through the Bussgang decomposition.
This allows it to maintain a recursive prediction and update structure even for quantized observations.
The architecture of the BKF is summarized in Figs.~\ref{fig:BussgangKalman} and \ref{fig:BKF}.

\subsection{Reduced-BKF for Multi-Branch 1-Bit Sensing} \label{section:rBKF}

Building upon the BKF, we also develop the reduced-BKF (rBKF) for multi-branch 1-bit sensing.
% As defined in Section~\ref{section:1-bit_state_estimation}, let $n_0$ denote the number of physical measurement features and let $L$ denote the number of parallel 1-bit comparator branches per feature.
% The aggregate binary observation dimension is then $n=L n_0$.
In the full BKF, the observation covariance matrix $\mathbf{S}_{t|t-1}\in\mathbb{R}^{L n_0 \times L n_0}$ must be inverted at each time step, and the dominant inversion cost scales cubically with $L n_0$.
The key idea of rBKF is to aggregate the comparator branches associated with the same physical feature before the gain update.
Let
\begin{equation}\begin{aligned}
    \mathbf{r}^*_t &= \mathbf{A}\mathbf{r}_t \\
    &= \mathbf{A}[\mathbf{r}_t[1,1],\ldots,\mathbf{r}_t[1,L], \mathbf{r}_t[2,1],\ldots,\mathbf{r}_t[n_0,L]]^{\top},
\end{aligned}\end{equation}
where $\mathbf{A}\in\mathbb{R}^{n_0\times Ln_0}$ is a branch-aggregation matrix and the uniform block-averaging projection is
\begin{align}
    \mathbf{A} = \mathbf{I}_{n_0} \otimes \frac{1}{L} \mathbf{1}_L^{\top}.
\end{align}
Equivalently,
\begin{align}
    \mathbf{r}_t^*[j] = \frac{1}{L} \sum_{\ell=1}^{L} \mathbf{r}_t[j,\ell],
    \qquad
    j=1, \ldots, n_0.
\end{align}
This projection averages only the comparator outputs corresponding to the same physical measurement feature.
% ; it does not average different state or measurement features.
Applying the projection to the Bussgang-linearized model in \eqref{eq:linearized_r} gives
\begin{align}
    &\mathbf{r}_t^* = \mathbf{A}\mathbf{r}_t
    = \mathbf{A}\mathbf{B}_t\mathbf{z}_t + \mathbf{A}\boldsymbol{\eta}_t
    = \mathbf{B}_t^*\mathbf{z}_t + \boldsymbol{\eta}_t^*, \\
    &\mathbf{B}_t^* = \mathbf{A}\mathbf{B}_t,
    \qquad
    \boldsymbol{\eta}_t^* = \mathbf{A}\boldsymbol{\eta}_t.
\end{align}
The reduced prior observation mean and covariance are
\begin{align}
    \hat{\mathbf{r}}_{t|t-1}^* = \mathbf{A}\hat{\mathbf{r}}_{t|t-1} \approx \mathbf{0},
    \qquad
    \mathbf{S}_{t|t-1}^* = \mathbf{A}\mathbf{S}_{t|t-1}\mathbf{A}^{\top}.
\end{align}

Based on these reduced-order moments, the rBKF gain is computed as
\begin{align}
    \mathbf{BG}_t^* = \mathbf{\Sigma}_{t|t-1} (\mathbf{B}_t^* \mathbf{H}_t)^{\top} (\mathbf{S}_{t|t-1}^*)^{-1}. \label{eq:rBG}
\end{align}
The posterior update is then
\begin{align}
    &\hat{\mathbf{x}}_{t|t} = \hat{\mathbf{x}}_{t|t-1} + \mathbf{BG}_t^* \left( \mathbf{r}_t^* - \hat{\mathbf{r}}_{t|t-1}^* \right), \\
    &\mathbf{\Sigma}_{t|t} = \mathbf{\Sigma}_{t|t-1} - \mathbf{BG}_t^* \mathbf{S}_{t|t-1}^* \mathbf{BG}_t^{*\top}.
\end{align}
Since $\hat{\mathbf{r}}_{t|t-1}^* \approx \mathbf{0}$ under the zero-centered construction, the correction term reduces to $\mathbf{BG}_t^* \mathbf{r}_t^*$.

When $L=1$, we have $n=n_0$ and $\mathbf{A}=\mathbf{I}_{n_0}$, so the rBKF reduces exactly to the BKF.
For $L>1$, the dominant matrix inversion is performed on $\mathbf{S}_{t|t-1}^* \in \mathbb{R}^{n_0 \times n_0}$ rather than on $\mathbf{S}_{t|t-1} \in \mathbb{R}^{Ln_0 \times Ln_0}$.
Therefore, the main cubic inversion cost is reduced from $\mathcal{O}((Ln_0)^3)$ to $\mathcal{O}(n_0^3)$, while retaining the prediction--correction structure of BKF.
% The numerical experiments in Section~\ref{section:numerical_exp} show that this reduced projection is nearly information-preserving under identical branch statistics and incurs only a small loss under heterogeneous branch statistics.

\begin{remark}[Projection matrix $\mathbf{A}$] \normalfont

    The rBKF projection averages only the $L$ binary outputs associated with the same physical feature.
    When the branches are statistically homogeneous, these outputs are exchangeable and the all-ones branch direction is sufficient for the Bussgang--LMMSE update; this explains why rBKF and BKF show nearly identical performance in the identical-noise setting.
    Under heterogeneous branch statistics, uniform averaging is no longer generally optimal, and reliability-weighted or learned projections may further improve performance.
    We use uniform averaging as a parameter-free low-complexity design and leave optimized projections for future work.
\end{remark}
% }}

%-------------------------------------------------------------------
\section{Bussgang-aided KalmanNet}
%-------------------------------------------------------------------

While BKF and rBKF are effective when the state-transition and measurement models are fully known, practical IoT sensing systems often suffer from partial model knowledge, model mismatches, and uncertain noise statistics.
% {\textcolor{blue}{
These effects can make the analytical Bussgang-Gaussian covariance model poorly calibrated.
% }}%
For this setting, we propose the Bussgang-aided KalmanNet (BKNet), a front-end-aware model-based deep learning estimator for 1-bit observations.
Rather than learning the entire filtering recursion in a purely end-to-end black-box manner, BKNet preserves the rBKF structure and learns only the reduced-order Bussgang gain from data.
This design preserves the recursive inductive bias of BKF and rBKF while keeping the model size manageable for resource-constrained implementations.

%-------------------------------------------------------------------
\subsection{High-Level Architecture}

We design BKNet based on the rBKF rather than the BKF.
This is because the reduced observation dimension in rBKF lowers both the computational overhead of the gain update and the size of the matrix-valued gain to be learned.
% Furthermore, as rBKF reduces to BKF when $a=1$ and $\mathbf{A} = \mathbf{I}$, the proposed architecture includes the full-dimensional case as a special case.
% {\textcolor{blue}{
% Furthermore, as rBKF reduces to BKF when $L=1$ and $\mathbf{A}=\mathbf{I}_{n_0}$, the proposed architecture includes the full-dimensional single-branch case as a special case.
Furthermore, when $L=1$, there is no branch aggregation to perform and $\mathbf{A}=\mathbf{I}_{n_0}$, so rBKF becomes identical to BKF.
Thus, although BKNet is built on the rBKF structure, it also covers the standard single-branch BKF case.
% }}

% {\color{blue}{
The state prediction, measurement prediction, adaptive threshold update, and branch aggregation follow the recursive structure of rBKF.
The only component replaced by learning is the analytical gain $\mathbf{BG}_t^*$ in \eqref{eq:rBG}.
Specifically, BKNet uses a learned gain mapping $\mathbf{BG}^{\text{net}}_t(\Theta)$, and the state update is given by
\begin{align}
    \hat{\mathbf{x}}_{t|t} = \hat{\mathbf{x}}_{t|t-1} + \mathbf{BG}^{\text{net}}_t(\Theta) \left(\mathbf{r}_t^* - \hat{\mathbf{r}}_{t|t-1}^* \right).
\end{align}
Under the zero-centered front-end construction, $\hat{\mathbf{r}}_{t|t-1}^* \approx \mathbf{0}$, and the correction term reduces approximately to $\mathbf{BG}^{\text{net}}_t(\Theta)\mathbf{r}_t^*$.
In this sense, BKNet maintains the same prediction--correction structure as rBKF while replacing only the analytical Bussgang gain with a data-driven gain.
% }}

% The state prediction, measurement prediction, and threshold update follow the recursive structure of the rBKF, whereas the analytically computed Bussgang gain $\mathbf{BG}^*_t$ is replaced by a learned mapping $\CMcal{BG}_t(\Theta)$.
% Accordingly, the state update of BKNet is given by:
% \begin{align}
%     \hat{\mathbf{x}}_{t|t} = \hat{\mathbf{x}}_{t|t-1} + \CMcal{BG}_t(\Theta) \bigl( \mathbf{r}^*_t -\hat{\mathbf{r}}^*_{t|t-1} \bigr).
% \end{align}

% Under the zero-centered front-end construction adopted in rBKF, $\hat{\mathbf{r}}^{*}_{t|t-1}=\mathbf{0}$, so the correction term reduces to $\CMcal{BG}_t(\Theta) \mathbf{r}^*_t$.
% In this sense, BKNet maintains the recursive correction structure of the rBKF while learning the gain in a data-driven manner.

This design follows the model-based deep learning paradigm \cite{mbdnn:book:23}.
It is also structurally related to KalmanNet \cite{kalmannet:tsp:22} in that both learn a correction gain rather than replacing the entire filtering pipeline.
The key difference lies in the observation model: while KalmanNet assumes high-resolution measurements, BKNet is specifically designed to handle zero-centered 1-bit observations generated by the sensing front-end.

%-------------------------------------------------------------------
\subsection{GRU-Based Gain Learning}

\begin{table*}[t]
    \centering
        % \caption{Input, Hidden State, and Output Definitions of the GRU Modules in BKNet}
        \caption{Input, hidden-state, and output representations of the GRU modules in BKNet. The covariance-related quantities indicate latent representations inspired by the corresponding analytical terms, rather than exact covariance estimates.}
    \begin{tabular}{|l|c|c|c|}
        \hline
         & GRU $\mathbf{Q}$ 
        & GRU $\mathbf{\Sigma}$ 
        & GRU $\mathbf{P}$ \\
        \hline \hline
        Input & $\text{FC}(\Delta \tilde{\mathbf{x}}_{t-1};{\Theta})$
        & $[\text{FC} \bigl( [\text{vec}(\mathbf{Q}_{t});\Delta \hat{\mathbf{x}}_{t-1}];{\Theta} \bigr);\text{vec}(\mathbf{Q}_{t})]$
        & $\bigl[ \text{FC}\bigl( \text{vec}(\mathbf{\Sigma}_{t|t-1});{\Theta} \bigr); \text{FC}\bigl( [\Delta \mathbf{r}_{t}; \Delta \hat{\mathbf{r}}_{t};{\Theta}] \bigr) \bigr]$ \\
        \hline
        Hidden state  & $\text{vec}(\mathbf{Q}_{t-1})$
        & $\text{vec}(\mathbf{\Sigma}_{t-1|t-1})$
        & $\text{vec}((\mathbf{B}^{*}_{t-1} \mathbf{H}_{t-1})^\top \cdot \mathbf{S}^{*\;-1}_{t-1|t-2})$ \\
        \hline
        Output & $\text{vec}(\mathbf{Q}_{t})$
        & $\text{vec}(\mathbf{\Sigma}_{t|t-1})$
        & $\text{vec}((\mathbf{B}^{*}_{t} \mathbf{H}_t)^\top \cdot \mathbf{S}^{*\;-1}_{t|t-1})$ \\
        \hline
    \end{tabular}
    \label{tab:GRU_BKN}
\end{table*}

The design of BKNet is motivated by the analytical form of the rBKF gain in \eqref{eq:rBG}, which depends on second-order quantities associated with the process noise, state uncertainty, and measurement uncertainty.
Instead of directly estimating these matrices, BKNet learns latent representations playing analogous roles in the gain computation.
To this end, BKNet employs three GRU modules, denoted by GRU $\mathbf{Q}$, GRU $\mathbf{\Sigma}$, and GRU $\mathbf{P}$, together with several fully connected (FC) layers.
The overall architecture is illustrated in Figs.~\ref{fig:BussgangKalman} and \ref{fig:BKN}.

For notational convenience, let $\text{FC}(\cdot;\Theta)$ denote an FC-based embedding layer with trainable parameters $\Theta$, let $[\mathbf{a};\mathbf{b}]$ denote vector concatenation, and let $\text{vec}(\cdot)$ denote matrix vectorization.
The input, hidden state, and output definitions of the GRU modules, together with the FC-based embeddings used in their inputs, are summarized in Table~\ref{tab:GRU_BKN}.
The roles of each module are summarized as follows.

\begin{enumerate}
    \item \textbf{GRU $\mathbf{Q}$}:
    The process-noise related representation is inferred from the temporal increments of the state estimates $\Delta \tilde{\mathbf{x}}_{t-1} = \hat{\mathbf{x}}_{t-1|t-1} - \hat{\mathbf{x}}_{t-2|t-2}$.
    This feature is embedded to dimension $m^2$ and fed into GRU $\mathbf{Q}$, whose hidden state stores the previous representation of the process-noise covariance.

    \item \textbf{GRU $\mathbf{\Sigma}$}:
    The prior state-error representation is learned using both the output of GRU $\mathbf{Q}$ and the innovation of the previous state estimate $\Delta \hat{\mathbf{x}}_{t-1} = \hat{\mathbf{x}}_{t-1|t-1} - \hat{\mathbf{x}}_{t-1|t-2}$.
    This reflects the analytical structure where $\mathbf{\Sigma}_{t|t-1}$ depends on the previous posterior covariance and the process-noise term.
    
    \item \textbf{GRU $\mathbf{P}$}:
    The measurement side representation is learned from the reduced 1-bit observation dynamics.
    In particular, we use $\Delta \mathbf{r}_t = \mathbf{r}^*_t - \mathbf{r}^*_{t-1}$ and $\Delta \hat{\mathbf{r}}_t = \mathbf{r}^*_t - \hat{\mathbf{r}}^*_{t|t-1}$ together with the projected prior state-error representation.
    The goal of this block is to learn a latent representation associated with $(\mathbf{B}^*_t \mathbf{H}_t)^{\top} \mathbf{S}^{*-1}_{t|t-1}$.

    \item \textbf{FC layers}:
    The outputs of GRU $\mathbf{\Sigma}$ and GRU $\mathbf{P}$ are combined through FC layers to generate the learned Bussgang gain $\mathbf{BG}^{\text{net}}_t(\Theta)$.
    Additional FC layers are used to update the posterior state-error representation, which is then fed back as the hidden state of GRU $\mathbf{\Sigma}$ at the next time step.
\end{enumerate}

By learning only the reduced-order Bussgang gain from zero-centered 1-bit observations, BKNet preserves the front-end-aware recursive structure of the rBKF while adapting to model mismatches through data-driven temporal representations.

%-------------------------------------------------------------------
\subsection{Training}

BKNet is trained end-to-end in a supervised manner using reduced 1-bit observation sequences and their corresponding ground-truth state sequences.
The loss for a single sequence of length $T$ is defined as the MSE:
\begin{align}
    \CMcal{L} = \frac{1}{T}\sum_{t=1}^T \left\|\mathbf{x}_t-\hat{\mathbf{x}}_{t|t}\right\|^2.
\end{align}
The training dataset consists of $N$ sequence pairs $\CMcal{D} = \{ (\mathbf{R}_i, \mathbf{X}_i) \}_{i=1}^N$, where
\begin{align}
    &\mathbf{X}_i = [\mathbf{x}^{(i)}_1,\mathbf{x}^{(i)}_2,\ldots,\mathbf{x}^{(i)}_{T_i}], \\
    &\mathbf{R}_i = [\mathbf{r}^{*(i)}_1,\mathbf{r}^{*(i)}_2,\ldots,\mathbf{r}^{*(i)}_{T_i}]. 
\end{align}
To mitigate overfitting, we incorporate $\ell_2$ regularization and use the sequence-level objective as:
\begin{align}
    \CMcal{L}^{(i)} = \frac{1}{T_i}\sum_{t=1}^{T_i} \left\| \mathbf{x}^{(i)}_t - \hat{\mathbf{x}}^{(i)}_{t|t} \right\|^2 + \lambda \| \Theta \|^2,
\end{align}
where $\lambda$ controls the balance between data fidelity and model complexity.
The mini-batch loss at iteration $k$ is given by:
\begin{align}
    \CMcal{L}^{\text{mini-batch}}_k = \frac{1}{|\CMcal{B}_k|}\sum_{j \in \CMcal{B}_k} \CMcal{L}^{(j)},
\end{align}
where $\CMcal{B}_k$ is the $k$-th mini-batch and $\| \CMcal{B}_k \| = B$ is the mini-batch size.
Since BKNet is differentiable with respect to $\Theta$, the parameters are optimized via standard backpropagation through time (BPTT) using Adam optimizer \cite{kingma2014adam}.
% {\textcolor{blue}{
The hard comparator and the applied 1-bit observations are treated as part of the sensing front-end; the optimization updates the gain-learning network rather than differentiating through the physical quantizer.
At test time, the learned parameters $\Theta$ are fixed; only the recurrent hidden states evolve along the sequence, and no gradient update or online retraining is performed.
Additional training and cross-validation curves for the nominal Lorenz and NCLT settings are provided in Appendix~\ref{app:learning_curves}.
% , where we discuss the optimization behavior and clarify that the held-out test trajectories are not used for gradient updates or checkpoint selection.
% }}

%-------------------------------------------------------------------
\section{Numerical Experiments} \label{section:numerical_exp}
%-------------------------------------------------------------------

In this section, we evaluate the proposed BKF, rBKF, and BKNet under both simulated and real-data 1-bit sensing settings.
% The evaluation is organized around the following aspects:
% i) can the proposed methods accurately reconstruct states from severely quantized observations,
% ii) how critical is the adaptive thresholding for zero-centering the front-end input,
% iii) what complexity and performance tradeoff is achieved by rBKF in multi-branch 1-bit sensing, and
% iv) how robust is the BKNet under partial model knowledge and real-world mobile sensing environment.
% {\textcolor{blue}{
% i) whether reliable state estimation is possible from single-branch 1-bit observations,
% ii) how the proposed front-end-aware methods compare with high-resolution estimators and quantization-aware references,
% iii) what accuracy--complexity tradeoff is achieved by rBKF in multi-branch 1-bit front-ends,
% iv) how BKNet behaves under partial model knowledge,
% v) whether the Bussgang--Gaussian covariance propagated by BKF is statistically calibrated,
% vi) how robust adaptive thresholding is under threshold mismatch and transient tracking errors, and
% vii) whether the BKF--BKNet performance gap on real NCLT trajectories can be explained by scalar covariance tuning.
% The evaluation focuses on four aspects: 1-bit state recovery, comparison with quantization-aware references, the rBKF accuracy–complexity tradeoff, and robustness under model, covariance, and threshold mismatch.
The evaluation focuses on four aspects: 1-bit state recovery, comparisons with representative baselines, the rBKF accuracy--complexity tradeoff, and robustness under model, covariance, and threshold mismatch.
The Lorenz attractor serves as a controlled benchmark with known dynamics, measurement model, and noise statistics, enabling multi-branch, model-mismatch, covariance-calibration, and threshold-robustness studies.
The Michigan NCLT dataset is used for real-trajectory validation under emulated 1-bit sensing, where exact branch-level front-end control and raw analog waveforms are unavailable.
% }}

% To this end, we use two complementary evaluation environments.
% The Lorenz attractor is used together with a simulated 1-bit comparator front-end, and serves as a synthetic benchmark in which the state-transition function $\mathbf{f}(\cdot)$, the measurement function $\mathbf{h}(\cdot)$, and the noise covariances $\mathbf{Q}$ and $\mathbf{R}$ are fully known and controllable.
% Accordingly, all controlled analyses, including multi-branch front-ends, branch-wise noise heterogeneity, reduced-order processing, and model mismatch, are conducted on the Lorenz attractor.
% In contrast, the Michigan NCLT dataset \cite{nclt} does not provide raw pre-ADC analog waveforms or exact branch-level front-end configurations, and is therefore not suitable for such fully controlled experiments.
% Instead, the NCLT dataset is used for real-world mobile sensing validation under emulated 1-bit sensing obtained by re-quantizing recorded measurements, and for verifying whether the benefit of adaptive thresholding carries over to real trajectory data.

As reference baselines, we employ the EKF \cite{schmidt1981kalman}, KalmanNet \cite{kalmannet:tsp:22}, and Split-KalmanNet \cite{split_kal:tvt:23} for single-branch configurations.
Their direct application to 1-bit observations should be interpreted not as quantizer-aware competitors, but rather as stress-test references that illustrate how estimators designed for high-resolution measurement fail under severe quantization.
% {\textcolor{blue}{
% To provide quantization-aware references, we additionally include SOI-KF \cite{SOI_KF} and an exact-likelihood particle filter (PF) \cite{Q_PF}.
% SOI-KF is a sign-of-innovation Kalman filter, i.e., a recursive 1-bit estimator that updates the state using the sign of the innovation rather than the analog innovation.
% For the Lorenz attractor, we adapt SOI-KF to the nonlinear setting by applying the same EKF-style Jacobian linearization used in the analytical filters.
% In the notation of BKF, SOI-KF can be interpreted as a diagonal-output-covariance approximation of BKF, i.e., $\mathbf{S}_{t|t-1} \approx \mathbf{I}$, which ignores the arcsine cross-correlations among simultaneous sign bits.
% As quantization-aware references, we include SOI-KF \cite{SOI_KF} and an exact-likelihood particle filter (PF) \cite{Q_PF}.
As quantization-aware references, we include the sign-of-innovations Kalman filter (SOI-KF) \cite{SOI_KF} and an exact-likelihood particle filter (PF) \cite{Q_PF}.
SOI-KF updates the state using the sign of the innovation and can be viewed as a diagonal sign-covariance approximation of BKF.
The PF uses the exact comparator likelihood and is reported on the controlled Lorenz benchmark as a likelihood-based accuracy reference.
For the exact-likelihood PF under the diagonal Lorenz measurement-noise setting, we use a bootstrap proposal with $N_p=1000$ particles and systematic resampling when the effective sample size falls below $N_p/2$.
The particle weights are updated using the comparator likelihood $p(r_t|x_t)=\prod_i \Phi(r_t[i](h_i(x_t)-\tau_t[i])/\sigma_i)$, using the same 1-bit observations and adaptive thresholds as BKF.
% }}

% {\textcolor{blue}{
% The exact-likelihood PF uses the comparator-induced 1-bit likelihood directly in the particle-weight update.
% For diagonal measurement noise, the component-wise likelihood is
% \begin{align}
%     p(\mathbf{r}_t | \mathbf{x}_t) = \prod_i \Phi\! \left( \frac{\mathbf{r}_t[i] \left(\mathbf{h}_i(\mathbf{x}_t) - \boldsymbol{\tau}_t[i] \right)}{ \sigma_i}
%     \right),
% \end{align}
% where $\Phi(\cdot)$ is the standard Gaussian CDF and $\sigma_i^2 = \mathbf{R}_t[i,i]$.
% We use $N_{\rm p}= 1000$ particles; increasing $N_{\rm p}$ beyond this value did not further reduce the MSE noticeably.
% The PF therefore serves as a near-Bayesian accuracy ceiling for the fully specified 1-bit observation model, at the cost of full model knowledge and higher computation.
% For fairness, BKF, SOI-KF, and PF use the same 1-bit observations and the same adaptive threshold rule $\boldsymbol{\tau}_t = \hat{\mathbf{y}}_{t|t-1}$; they differ only in the measurement-update rule.
% }}

Furthermore, Appendix~\ref{app:split_bknet} provides additional comparisons with Split-BKNet, which maintains the same adaptive-thresholding front-end and reduced observation path as BKNet while replacing the gain-learning backbone with a Split-KalmanNet architecture.
This comparison is deferred to the appendix because the primary focus of this paper is on sensing-front-end-aware design under 1-bit observations, whereas Split-BKNet serves as a heavier learning variant that, while potentially yielding higher accuracy, requires significantly larger model sizes and computational costs.

%-------------------------------------------------------------------
\subsection{Lorenz Attractor Setups and Evaluation Metrics}

The Lorenz dynamics are written in state-dependent coefficient form as
\begin{equation}\begin{aligned}
    \frac{d\mathbf{x}_{\tau}}{d\tau} = \mathbf{g}(\mathbf{x}_{\tau}) 
    = \mathbf{A}(\mathbf{x}_{\tau})\mathbf{x}_{\tau},
    = \begin{bmatrix} 
    -\sigma & \sigma & 0 \\
    \rho-x_{3,\tau} & -1 & 0\\
    x_{2,\tau} & 0 & -\beta \end{bmatrix}
    \mathbf{x}_{\tau},
    \label{eq:lorenz_continuous}
\end{aligned}\end{equation}
where
$\mathbf{x}_{\tau} = [x_{1,\tau},x_{2,\tau},x_{3,\tau}]^{\top}$ and $(\sigma,\rho,\beta)=(10,28,8/3)$.
Assuming that $\mathbf{A}(\mathbf{x}_{\tau})$ is locally constant over a sampling interval $\Delta\tau$, its order-$J$ discrete approximation is
\begin{align}
    \boldsymbol{\Phi}_{J}(\mathbf{x}_{\tau}) \triangleq \mathbf{I}_{3} + \sum_{j=1}^{J} \frac{\left( \mathbf{A}(\mathbf{x}_{\tau})\Delta\tau \right)^{j}}{j!}. \label{eq:lorenz_transition_operator}
\end{align}
Defining $\mathbf{f}_{J}(\mathbf{x}) \triangleq \boldsymbol{\Phi}_{J}(\mathbf{x})\mathbf{x}$, the discrete-time model is
\begin{align}
    \mathbf{x}_{t+1} = \mathbf{f}_{J}(\mathbf{x}_{t}) + \mathbf{w}_{t+1}. \label{eq:lorenz_discrete_model}
\end{align}
Unless otherwise stated, $\Delta\tau=0.02$ and $J=5$.

For the Lorenz experiments, the state dimension and the underlying physical measurement-feature dimension are both $m=n_0=3$.
In the single-branch case, we set $L=1$, $n=n_0$, and $\tilde{\mathbf{h}}(\mathbf{x}_t)=\mathbf{x}_t$.
In the multi-branch case, $L$ parallel 1-bit comparator branches are used per physical feature, so the aggregate binary observation dimension becomes $n=Ln_0$, following the front-end model in Section~\ref{section:1-bit_state_estimation}.
% Unless otherwise specified, the process and measurement noise covariances are written as $\mathbf{Q}=q^2 \mathbf{I}_m$ and $\mathbf{R}=r^2 \mathbf{I}_n$, respectively, with the specific values of $q^2$ and $r^2$ given in each experiment.
Unless otherwise specified, the nominal Lorenz noise setting is $q^2=10^{-3}$ and $r^2=10^{-1}$, i.e., $\mathbf{Q}=q^2 \mathbf{I}_m$ and $\mathbf{R}=r^2 \mathbf{I}_n$.
% }}%
% For the single-branch case, we set $\tilde{\mathbf{h}}(\mathbf{x}_t) = \mathbf{x}_t$ as the measurement function.
% For the multi-branch case, the measurement function is extended according to the configuration.
The training data consists of paired sequences $\{ (\mathbf{Y}_i, \mathbf{X}_i) \}^N_{i=1}$, and the quantized observation sequence $\mathbf{R}_i$ is generated by the 1-bit front-end with threshold update as in Fig.~\ref{fig:BussgangKalman_architecture}.
The sequence length is set to 100 for training and 2000 for testing.
We use MSE in dB scale as the main accuracy metric:
\begin{align}
    \text{MSE\,[dB]} = 10\log_{10} \! \left( \frac{1}{\sum_i T_i} \sum_{i=1}^N \sum_{t=1}^{T_i} \| \mathbf{x}_t - \hat{\mathbf{x}}_{t|t} \|^2 \right).
\end{align}
For the rBKF, we also measure the inference time to assess the complexity benefit of reduced-order processing.

% {\color{blue}{
For the analytical filters that propagate posterior covariance, especially BKF and rBKF, we additionally evaluate covariance calibration.
Let $\mathbf{e}_t=\mathbf{x}_t-\hat{\mathbf{x}}_{t|t}$ denote the posterior estimation error and let $\mathbf{\Sigma}_{t|t}$ denote the posterior covariance propagated by the filter.
Over an evaluation index set $\mathcal{T}$, we compute the average normalized estimation error squared (ANEES) as
\begin{align}
    \text{ANEES} = \frac{1}{|\mathcal{T}|} \sum_{t\in\mathcal T} \mathbf{e}_t^\top \left( \mathbf{\Sigma}_{t|t} + \epsilon_{\text{reg}} \mathbf{I}_m \right)^{-1} \mathbf{e}_t,
\end{align}
where $\epsilon_{\text{reg}}>0$ is used only for numerical stabilization.
Under this normalization, a value close to $m$ indicates a well-calibrated covariance estimate.
We also report the trace ratio
\begin{align}
    \text{TR} = \frac{\sum_{t \in \mathcal{T}} \|\mathbf{e}_t\|_2^2} {\sum_{t \in \mathcal{T}} \mathrm{tr}(\mathbf{\Sigma}_{t|t})},
\end{align}
where values larger than one indicate overconfidence and values smaller than one indicate conservative covariance estimates.
In addition, when empirical error covariance estimates are available, we compute the normalized covariance error (NCE)
\begin{align}
    \text{NCE} = \frac{\|\hat{\mathbf{C}}_\mathbf{e} - \bar{\mathbf{\Sigma}} \|_F}{\|\bar{\mathbf{\Sigma}}\|_F},
    \quad
    \hat{\mathbf{C}}_\mathbf{e} = \frac{1}{|\mathcal{T}|} \sum_{t \in \mathcal{T}} \mathbf{e}_t \mathbf{e}_t^\top,
    \quad
    \bar{\mathbf{\Sigma}} = \frac{1}{|\mathcal{T}|} \sum_{t \in \mathcal{T}} \mathbf{\Sigma}_{t|t}.
\end{align}
The evaluation set $\mathcal{T}$ is specified in each diagnostic experiment; unless otherwise stated, it contains all test time instants.

For the zero-centering and Gaussian-surrogate diagnostics, we further report the normalized mean mismatch $\epsilon_{\mu_\mathbf{z}}$ and covariance mismatch $\epsilon_\mathbf{P}$ of the thresholded quantizer input $\mathbf{z}_t = \mathbf{y}_t - \boldsymbol{\tau}_t$:
\begin{align}
    \epsilon_{\mu_\mathbf{z}} = \frac{1}{|\mathcal{T}|} \sum_{t \in \mathcal{T}} \frac{\|\hat{\mu}_{\mathbf{z},t}\|_2} {\sqrt{\mathrm{tr}(\mathbf{P}_{t|t-1})}},
    \quad
    \epsilon_\mathbf{P} = \frac{1}{|\mathcal{T}|} \sum_{t \in \mathcal{T}} \frac{\|\hat{C}_{\mathbf{z},t} - \mathbf{P}_{t|t-1}\|_F} {\|\mathbf{P}_{t|t-1}\|_F}.
\end{align}
Here, $\hat{\mu}_{\mathbf{z},t}$ and $\hat{\mathbf{C}}_{\mathbf{z},t}$ denote empirical estimates of the mean and covariance of $\mathbf{z}_t$, respectively.
These metrics quantify how closely the adaptive thresholding mechanism realizes the approximately zero-centered Gaussian prior assumed by the BKF.

% Finally, for threshold-robustness and recovery experiments, we use the sign imbalance
% \begin{align}
%     \text{SI} = \frac{1}{n} \sum_{i=1}^{n} \left| \frac{1}{|\mathcal{T}|} \sum_{t \in \mathcal{T}} \mathbf{r}_t[i] \right|,
% \end{align}
% where a value close to zero indicates that the binary observations do not collapse to one sign.
% We also report the recovery time after an injected perturbation, defined as the number of time steps required for the instantaneous estimation error to return to the nominal steady-state error band and remain within that band for a prescribed window.
% Unless otherwise stated, the diagnostic metrics are computed after discarding the first $T_{\text{b}}=100$ time steps as burn-in, in order to remove initialization transients.
% The MSE values are reported over the full test sequence for consistency with the other performance tables.
Finally, for threshold-robustness and recovery experiments, we use the sign imbalance
\begin{align}
    \text{SI} = \frac{1}{n} \sum_{i=1}^{n} \left| \frac{1}{|\mathcal{T}|} \sum_{t \in \mathcal{T}} \mathbf{r}_t[i] \right|
\end{align}
where a value close to zero indicates that the binary observations do not collapse to one sign.
For state-perturbation experiments, we also report the transient error and the recovery time.
The transient error is the MSE in dB computed over the post-perturbation transient window $\mathcal T_{\mathrm{tr}}$,
\begin{align}
    E_{\mathrm{tr}}[\text{dB}] = 10\log_{10} \left( \frac{1}{|\mathcal T_{\mathrm{tr}}|} \sum_{t\in\mathcal T_{\mathrm{tr}}} \|\mathbf x_t-\hat{\mathbf x}_{t|t}\|^2 \right).
\end{align}
The recovery time is defined as the number of post-perturbation steps until the $20$-step moving-average squared error first falls below a threshold $3~\mathrm{dB}$ above the nominal steady-state MSE and remains below it for $20$ consecutive steps; $\mathcal{T}_{\mathrm{tr}}$ comprises the first $100$ post-perturbation steps.
Except for these perturbation-centered metrics, diagnostic metrics are computed after discarding the first $T_b=100$ time steps as burn-in, whereas MSE values are reported over the full test sequence.

% The recovery time is defined as the number of time steps required for the instantaneous estimation error to return to the nominal steady-state error band and remain within that band for a prescribed window.
% Unless otherwise stated, diagnostic metrics are computed after discarding the first $T_b=100$ time steps as burn-in, while MSE values are reported over the full test sequence for consistency with the other performance tables.

% For covariance calibration, we report ANEES, trace ratio (TR), and normalized covariance error (NCE).
% ANEES close to one indicates calibrated covariance, TR larger than one indicates overconfidence in the covariance trace, and NCE compares the empirical error covariance with the propagated covariance.
% We also report the normalized mean/covariance mismatch of the thresholded input $\mathbf{z}_t$ and the recovery time after injected perturbations.
% }}

%-------------------------------------------------------------------
\subsection{Single-Branch 1-Bit State Estimation}

\begin{figure*}[t]
    \centering
    \includegraphics[width=1.0\textwidth]{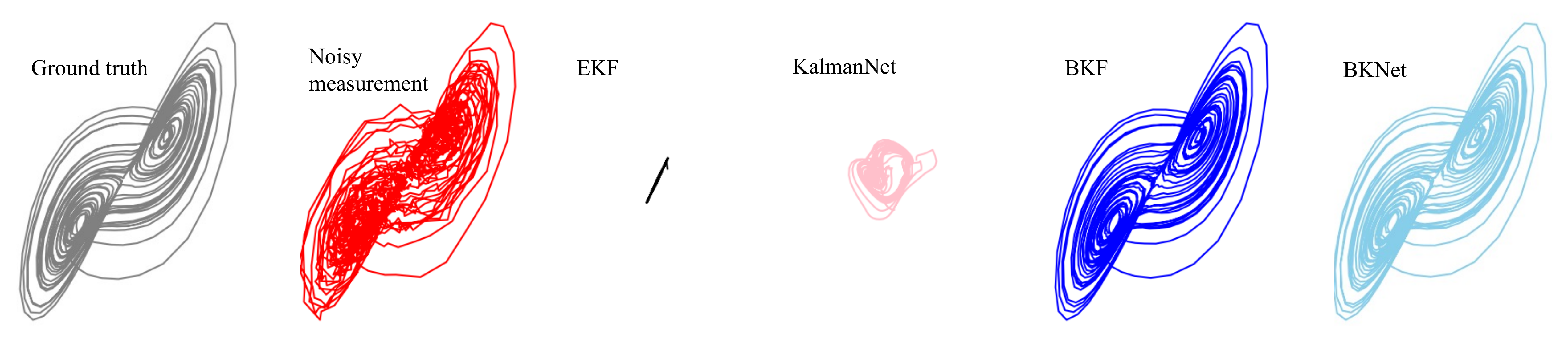}
    \vspace{-2em}
    \caption{State estimation results for the Lorenz attractor using 1-bit observations.}
    \label{fig:Lorenz_BKF_BKN}
\end{figure*}

First, we consider the single-branch case where only a single 1-bit comparator is used per measurement feature of the Lorenz attractor.
As references, EKF, KalmanNet, and Split-KalmanNet are evaluated both with ideal observations and direct 1-bit observations.
The latter serves as a stress-test reference under severe quantization.
The results are summarized in Table~\ref{tab:Lorenz_BKF_BKN}.

\begin{table}[t]
    \centering
    \caption{Numerical MSE [dB] for Lorenz attractor}
    \begin{tabular}{|c||c|c|}
        \hline
        Method                              & 1-bit observation      & Ideal observation  \\ \hline \hline
        EKF                                 & 17.85                  & -19.31             \\ \hline
        KalmanNet                           & 12.95                  & -19.49             \\ \hline
        Split-KalmanNet                     & 11.37                  & -18.93             \\ \hline
        SOI-KF                              & -16.29                 & --                 \\ \hline
        BKF                                 & -17.38                 & --                 \\ \hline
        BKNet                               & -17.31                 & --                 \\ \hline
        Exact-likelihood PF                 & -18.61                 & --                 \\ \hline
    \end{tabular}
    \label{tab:Lorenz_BKF_BKN}
\end{table}

As shown in Table~\ref{tab:Lorenz_BKF_BKN}, directly applying EKF, KalmanNet, or Split-KalmanNet to 1-bit observations fails to provide meaningful state estimates,
% {\textcolor{blue}{
confirming that high-resolution observation updates are fundamentally mismatched to the comparator front-end.
Among the quantization-aware recursive filters, SOI-KF already achieves a large improvement over these stress-test baselines, but it remains about $1.1\;\text{dB}$ worse than BKF.
This gap is consistent with the fact that SOI-KF ignores the simultaneous sign-bit correlations captured by the arcsine covariance in BKF.
% The exact-likelihood PF achieves the lowest MSE among the 1-bit methods, as expected from its direct use of the comparator likelihood; however, it requires the full likelihood model and has a higher inference time of 2.86 s, compared with 0.54 s for SOI-KF.
% The exact-likelihood PF achieves the lowest MSE among the 1-bit methods, as expected from its direct use of the comparator likelihood.
% However, it requires the full likelihood model and particle propagation, so we use it as a likelihood-based accuracy reference rather than as a low-complexity recursive baseline.
% These results show that BKF closes most of the gap to a likelihood-based Bayesian reference while retaining a lightweight recursive update.
The exact-likelihood PF achieves the lowest MSE among the 1-bit methods, as expected from its direct use of the fully specified comparator likelihood and particle-based posterior approximation.
However, this accuracy requires full model knowledge and substantially higher computation: in our implementation, the inference times of SOI-KF, BKF, BKNet, and PF are $0.540\;\text{s}$, $0.691\;\text{s}$, $1.032\;\text{s}$, and $2.859\;\text{s}$, respectively.
BKF therefore provides a favorable accuracy--complexity tradeoff between SOI-KF and PF: it improves over SOI-KF by $1.09\;\text{dB}$, remains only $1.23\;\text{dB}$ behind PF, and is approximately $4.1\times$ faster than PF.
BKNet likewise remains only $1.30\;\text{dB}$ behind PF while being approximately $2.8\times$ faster, and its learned-gain formulation is intended for partial-model settings in which the exact transition or comparator-likelihood model may be unavailable or unreliable.
Thus, we use the exact-likelihood PF as a near-Bayesian accuracy reference for the fully specified Lorenz model, whereas BKF and BKNet provide substantially lower-cost recursive alternatives for fully known and partially known models, respectively.
\subsection{Multi-Branch Front-End and Reduced-BKF} \label{section:exp_rBKF}

All results reported below, prior to Section~\ref{section:exp_NCLT}, are conducted using the Lorenz attractor.
This is because the following experiments require a controllable multi-branch 1-bit front-end and full model knowledge of the state-transition function $\mathbf{f}(\cdot)$, the measurement function $\mathbf{h}(\cdot)$, and the noise covariances $\mathbf{Q}$ and $\mathbf{R}$.

% We evaluate the BKF and rBKF when multiple parallel 1-bit comparator branches are used per measurement feature.
% The number of comparator branches is set to $1$, $8$, $64$ and $128$.
% {\color{blue}{
We evaluate BKF and rBKF when $L\in\{1,8,64,128\}$ parallel 1-bit comparator branches are used per physical measurement feature.
% For the Lorenz attractor, the physical measurement-feature dimension is $n_0=3$, and the aggregate binary observation dimension is therefore $n=Ln_0$.
% }}%
We consider two noise settings:
i) identical-noise branches, where all comparator branches have independent and identically distributed (i.i.d.) measurement noise with the same variance, and
ii) heterogeneous-noise branches, where the measurement noise variance varies across branches.

\subsubsection{Identical-noise branches}

\begin{figure}[t]
    \centering
    \includegraphics[width=0.5\textwidth]{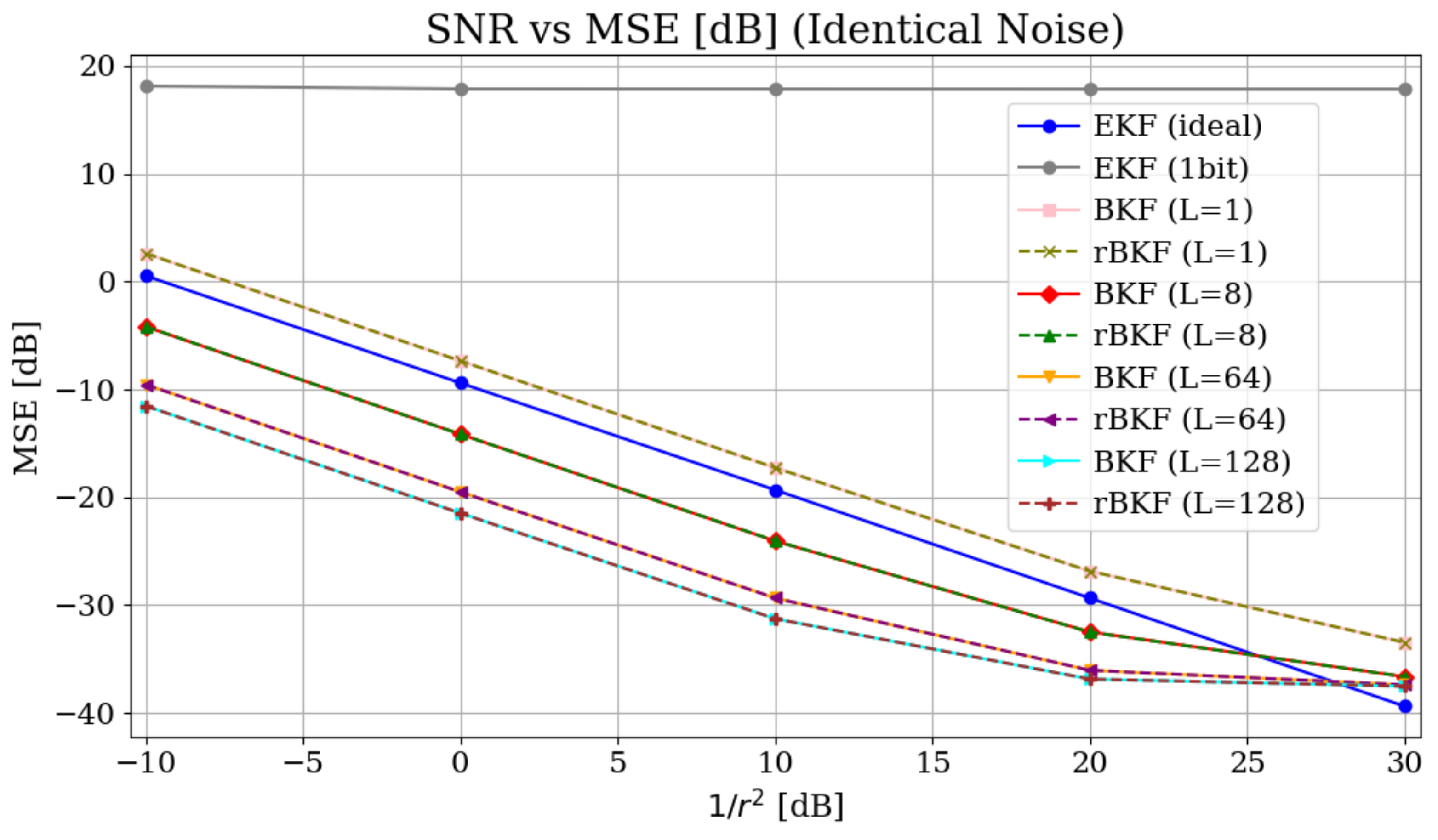}
    \vspace{-2em}
    \caption{The MSE performance comparison of BKF and rBKF with identical-noise.
    For $L=1$, BKF and rBKF coincide, so the corresponding curves overlap.}
    \label{fig:reduced-BKF_identical_MSE}
\end{figure}

In the first setting, the measurement noise covariance is set to $\mathbf{R} = r^2 \mathbf{I}$ with the signal-to-noise ratio (SNR) fixed at $-20\;\text{dB}$.
We vary $1/r^2$ from $-10\;\text{dB}$ to $30\;\text{dB}$.
Fig.~\ref{fig:reduced-BKF_identical_MSE} shows that BKF and rBKF exhibit nearly identical performance.
This implies that under statistically symmetric branch conditions, the averaging-based projection in rBKF preserves most of the useful information in the binary observations.

% The corresponding inference times are presented in Table~\ref{tab:reduced-BKF_identical_interference_time}.
% While the rBKF incurs a slight overhead in the single-branch case due to the projection step, its advantage grows rapidly as the number of branches increases.
% {\color{blue}{
The corresponding inference times are presented in Table~\ref{tab:reduced-BKF_identical_interference_time}. The EKF reference uses a single high-resolution observation and therefore does not scale with $L$.
While rBKF incurs a slight overhead when $L=1$, its advantage appears from $L=8$ and grows rapidly as $L$ increases.
% }}%
This confirms the benefit of reduced-order processing for resource-constrained implementations.
In some multi-branch settings, BKF and rBKF even outperform the single-branch ideal-observation EKF reference.
This should be interpreted as a benefit of front-end branch diversity rather than as a general claim that 1-bit sensing is intrinsically superior to high-resolution state estimation.

\begin{table}[t]
    \centering
        \caption{BKF $\&$ rBKF, $\#$ of comparator branches $L$ vs Inference time [s], identical-noise}
    \begin{tabular}{|c|c|c|c|c|}
        \hline
        $L$  & 1        & 8        & 64        & 128       \\ \hline \hline
        EKF (ideal observation, $L=1$)    & \multicolumn{4}{c|}{0.601}    \\ \hline
        BKF                  & 0.691    & 0.777    & 2.746     & 5.428     \\ \hline
        rBKF                 & 0.720    & 0.732    & 0.740     & 0.894     \\ \hline
    \end{tabular}
    \label{tab:reduced-BKF_identical_interference_time}
\end{table}

% {\color{blue}{
We emphasize that Table~\ref{tab:reduced-BKF_identical_interference_time} reports estimator-side inference time rather than analog front-end conversion latency.
% The front-end latency is governed by comparator decision time and programmable-reference settling, as discussed in Section~\ref{section:energy}.
Nevertheless, Table~\ref{tab:reduced-BKF_identical_interference_time} is relevant to resource-constrained deployment because it shows that the proposed reduced-order processing suppresses the digital filtering latency induced by increasing the number of 1-bit branches.
For example, when $L=128$, the inference time decreases from $5.428\;\text{s}$ for BKF to $0.894\;\text{s}$ for rBKF.
% }}

\subsubsection{Heterogeneous-noise branches}

To reflect more realistic front-end conditions, we also evaluate heterogeneous-noise branches.
The state noise variance is set to $q^2=-30\;\text{dB}$, and the measurement noise variance for each branch is independently sampled from a uniform distribution between $-20\;\text{dB}$ and $-10\;\text{dB}$.
The results are summarized in Table~\ref{tab:reduced-BKF_heterogeneous_MSE}.

\begin{table}[t]
    \centering
        \caption{BKF $\&$ rBKF, $\#$ of comparator branches $L$ vs MSE [dB], heterogeneous noise}
    \label{tab:reduced-BKF_heterogeneous_MSE}
    \begin{tabular}{|c|c|c|c|c|}
        \hline
        $L$                  & 1         & 8         & 64         & 128        \\ \hline \hline
        EKF (ideal observation, L=1)      & \multicolumn{4}{c|}{-22.41}     \\ \hline
        BKF                  & -20.53    & -26.59    & -31.80     & -33.79     \\ \hline
        rBKF                 & -20.53    & -26.33    & -31.45     & -33.33     \\ \hline
    \end{tabular}
\end{table}

As shown in Table~\ref{tab:reduced-BKF_heterogeneous_MSE}, both BKF and rBKF show performance improvements as the number of comparator branches increases.
Compared with the BKF, the rBKF incurs only a small loss under heterogeneous-noise conditions, while offering a large reduction in computational cost.
Overall, the results indicate that multi-branch 1-bit sensing can effectively mitigate quantization distortion, with rBKF offering a favorable complexity and performance tradeoff.

%-------------------------------------------------------------------
\subsection{BKNet under Partial Model Knowledge}

We now evaluate the performance of BKNet when the process and measurement statistics are only partially known.
The branch and noise configurations are kept consistent with the previous BKF/rBKF experiments.

Under identical-noise conditions, BKNet shows a steady improvement in performance as the number of comparator branches increases, similar to the trend observed in BKF and rBKF.
% The inference time also shows only a slight increase, remaining in the range of approximately 1.03--1.05 seconds; therefore, we omit a separate table for brevity.
% Instead, the detailed MSE and inference-time results under identical-noise conditions are provided in Appendix~\ref{app:bknet}.
The inference time also shows only a slight increase, remaining in the range of approximately $1.03$--$1.05$ seconds.
Since these results follow the same qualitative trend as BKF and rBKF, we provide the detailed identical-noise BKNet results in Appendix~\ref{app:bknet}.

Under heterogeneous-noise conditions, the performance of BKNet is summarized in Table~\ref{tab:BKN_heterogeneous_MSE}.
\begin{table}[t]
    \centering
    \caption{BKNet, $\#$ of comparator branches $L$ vs MSE [dB], heterogeneous noise}
    \begin{tabular}{|c|c|c|c|c|}
        \hline
        $L$                 & 1         & 8         & 64         & 128        \\ \hline \hline
        KalmanNet (ideal obs, L=1)           & \multicolumn{4}{c|}{-22.67}                     \\ \hline
        BKNet               & -20.83    & -26.55    & -31.96     & -34.05     \\ \hline
    \end{tabular}
    \label{tab:BKN_heterogeneous_MSE}
\end{table}
Even when the measurement side uncertainty is heterogeneous, the gains derived from increasing the number of comparator branches are clearly maintained.
As with the rBKF, these results should be interpreted as the benefits of front-end branch diversity and reduced-order processing rather than a universal superiority of 1-bit sensing over all high-resolution configurations.

Next, we evaluate BKNet under representative model-mismatch scenarios.
To keep the evaluation concise, we consider one dynamics-mismatch scenario and one measurement-model-mismatch scenario. 
For the dynamics mismatch, the true trajectories are generated using the discretization order $J=5$, whereas the estimator assumes the mismatched setting $J=1$. 
For the measurement-model mismatch, we rotate the measurement function $\mathbf{h}(\cdot)$ by $3^\circ$ with respect to the nominal model.
The results are summarized in Table~\ref{tab:BKN_model_mismatch}.

\begin{table}[t]
    \centering
    \caption{Representative model-mismatch results [dB]}
    \begin{tabular}{|c|c|c|c|}
        \hline
                  & Nominal & $J=1$         & $\mathbf{h}(\cdot)$ rotation  \\ \hline \hline
        BKF       & -17.38                  & 6.24      & -11.66                        \\ \hline
        BKNet     & -17.31                  & -15.19    & -14.20                        \\ \hline
    \end{tabular}
    \label{tab:BKN_model_mismatch}
\end{table}

BKNet exhibits significantly more robust performance than BKF under both mismatch conditions.
This indicates that learning only the front-end-aware correction gain is sufficient to compensate for model errors in 1-bit observation environments.
For a more exhaustive analysis, Appendix~\ref{app:split_bknet} provides additional comparisons using a Split-KalmanNet-based gain learner within the same front-end-aware framework.

%-------------------------------------------------------------------
% {\color{blue}{
\subsection{Bussgang--Gaussian and Filter-Consistency Diagnostics} \label{section:exp_consistency}

% \begin{figure}[t]
%     \centering
%     \includegraphics[width=0.45\textwidth]{image_rev/histogram.pdf}
%     \vspace{-0.5em}
%     \caption{{\color{blue}{Whitened prior innovation histograms of BKF under the nominal Lorenz setting.}}}
%     \vspace{-0.5em}
%     \label{fig:histogram}
% \end{figure}

\begin{table}[t]
    \centering
    \caption{Representative BKF covariance-calibration diagnostics under nominal and mismatched covariance settings on the Lorenz attractor.}
    % {\color{blue}{
    \begin{tabular}{|l||c|c|c|c|c|}
        \hline
        Case                                & $\alpha$ & MSE [dB] & TR      & NCE   & ANEES   \\ \hline\hline
        Nominal                             & 1        & -17.38   & 0.998   & 0.313 & 4.71    \\ \hline
        $\bar{\mathbf{Q}}=\alpha\mathbf{Q}$ & 0.01     & -16.64   & 1.966   & 0.475 & 244.56  \\ \hline
        $\bar{\mathbf{Q}}=\alpha\mathbf{Q}$ & 100      & -11.02   & 0.652   & 0.521 & 1.92    \\ \hline
        $\bar{\mathbf{R}}=\alpha\mathbf{R}$ & 0.01     & 2.65     & 1531.59 & 0.999 & 4117.52 \\ \hline
        $\bar{\mathbf{R}}=\alpha\mathbf{R}$ & 100      & -6.99    & 0.182   & 4.536 & 5.65    \\ \hline
    \end{tabular}%}}
    \label{tab:covariance_mismatch}
    \vspace{-0.5em}
\end{table}

We examine the calibration of the covariance propagated by BKF under the zero-centered Bussgang--Gaussian surrogate.
Since BKF does not propagate the exact posterior covariance of the nonlinear 1-bit observation model, these diagnostics are intended to quantify the reliability of the propagated second-order approximation rather than to prove exact posterior Gaussianity.

In the nominal Lorenz tracking regime, the normalized mean and covariance mismatches of the thresholded quantizer input are $\epsilon_{\mu_\mathbf{z}}=0.0410$ and $\epsilon_P=0.1238$, respectively.
This indicates that adaptive thresholding approximately zero-centers the quantizer input in the tested tracking regime.
Table~\ref{tab:covariance_mismatch} further evaluates the posterior covariance propagated by BKF.
In the nominal case, the trace ratio is close to one, $\mathrm{TR}=0.998$, suggesting that the overall covariance scale is reasonable.
However, $\mathrm{ANEES}=4.71$ remains larger than $m=3$, indicating mild-to-moderate overconfidence in the Mahalanobis sense.

We next test covariance sensitivity by running BKF with mismatched covariance parameters.
For process-covariance mismatch, the filter uses $\bar{\mathbf{Q}}=\alpha\mathbf{Q}$ while $\mathbf{R}$ is fixed; for measurement-covariance mismatch, the filter uses $\bar{\mathbf{R}}=\alpha\mathbf{R}$ while $\mathbf{Q}$ is fixed.
Representative under- and over-estimated cases are summarized in Table~\ref{tab:covariance_mismatch}.
Underestimating $\mathbf{Q}$ or $\mathbf{R}$ makes BKF overconfident, as reflected by the increased TR and ANEES, with the most severe degradation occurring when $\bar{\mathbf{R}} = 0.01 \mathbf{R}$.
Overestimating the covariances also degrades MSE and NCE, indicating inaccurate or overly conservative covariance propagation.
These results show that BKF provides a useful analytical recursive approximation when the assumed statistics are close to the data-generating model, whereas severe covariance mismatch can lead to poorly calibrated uncertainty.
This motivates the learned-gain adaptation of BKNet under partial model knowledge.

\subsection{Robustness of Adaptive Thresholding}

% The controlled perturbation in \eqref{eq:threshold_mismatch} can also be viewed as a simplified model of practical comparator-reference mismatch.
% It does not cover all circuit-level impairments, but it provides a first-order stress test for the most relevant nonideality in the proposed adaptive-threshold front-end: imperfect alignment between the predicted measurement and the actual comparator reference.

% Adaptive thresholding is designed to make the quantizer input approximately zero-centered, but this condition depends on the quality of the one-step prediction.
% We therefore evaluate not only the benefit of threshold update in the nominal tracking regime, but also its robustness under controlled threshold mismatch and transient state-estimation errors.
% In this subsection, BKNet-Origin denotes the BKNet trained under the nominal setting, whereas BKNet-Aug denotes the same BKNet architecture trained with threshold and transient-perturbation augmentation.
We evaluate the robustness of the adaptive-threshold front-end under two representative failure modes: systematic threshold mismatch and transient state-estimation errors.
BKNet-Origin denotes the BKNet trained under the nominal setting, whereas BKNet-Aug denotes the same architecture trained with threshold and transient-perturbation augmentation.

% We first isolate the effect of the adaptive-threshold update by keeping the BKNet backbone fixed and removing only the threshold update.
% Table~\ref{tab:threshold_ablation_bknet} shows that adaptive thresholding substantially improves the MSE on both the Lorenz attractor and the NCLT dataset.
% The NCLT setup is described in Section~\ref{section:exp_NCLT}.
% This confirms that adaptive thresholding is not merely a mathematical simplification for the zero-mean Bussgang model, but a practical front-end mechanism for preserving informative sign variations in 1-bit sensing.

% \begin{figure}[t]
%     \centering
%     \includegraphics[width=0.45\textwidth]{image_rev/threshold_mismatch_MSE.pdf}
%     \label{fig:threshold_mismatch_mse}
%     \caption{{\color{blue}{Robustness of MSE to controlled threshold mismatch on the Lorenz attractor.}}}
% \end{figure}

% \begin{figure}
%     \centering
%     \includegraphics[width=0.45\textwidth]{image_rev/threshold_mismatch_eps.pdf}
%     \label{fig:threshold_mismatch_eps}
%     \caption{{\color{blue}{Robustness of $\epsilon_{\mu_\mathbf{z}}$ to controlled threshold mismatch on the Lorenz attractor.}}}
% \end{figure}

First, adaptive thresholding itself provides a large performance gain.
With the BKNet architecture otherwise unchanged, replacing the adaptive threshold with a fixed zero threshold degrades the MSE from $-17.31$ to $-14.26$ dB on the Lorenz attractor.
On the NCLT dataset, however, training with the fixed zero threshold diverges, and thus no meaningful test MSE can be reported.
% With the BKNet backbone fixed, replacing the adaptive threshold with a fixed zero threshold degrades the MSE from $-17.31$ to $-14.26\;\text{dB}$ on the Lorenz attractor and from $18.62$ to $135.4\;\text{dB}$ on the NCLT dataset.
% This confirms that the threshold update is not only a mathematical device for simplifying the Bussgang model, but also a practical mechanism for preserving informative sign variations in 1-bit sensing.
This confirms that the threshold update is not only a mathematical device for simplifying the Bussgang model, but also an algorithmically useful mechanism for preserving informative sign variations in 1-bit sensing.

\begin{table}[t]
    \centering
    \caption{Representative threshold-mismatch results on the Lorenz attractor.}
    \label{tab:threshold_mismatch}
    % {\color{blue}{
    \begin{tabular}{|c||c|c|c|}
    \hline
    $\delta_\tau$ & Method & MSE [dB] & $\epsilon_{\mu_\mathbf{z}}$ \\
    \hline\hline
    \multirow{2}{*}{0}
    & BKF          & -17.38 & 0.041 \\
    & BKNet-Origin & -17.31 & 0.124 \\
    \hline
    \multirow{3}{*}{0.5}
    & BKF          & -16.00 & 0.597 \\
    & BKNet-Origin & -15.43 & 0.602 \\
    & BKNet-Aug    & -16.77 & 0.612 \\
    \hline
    \multirow{3}{*}{1.0}
    & BKF          & -12.99 & 1.060 \\
    & BKNet-Origin &  -9.19 & 1.123 \\
    & BKNet-Aug    & -15.17 & 1.090 \\
    \hline
    \end{tabular}%}}
\end{table}

To test whether the zero-centering condition is fragile, we deliberately perturb the adaptive threshold as
\begin{align}
    \boldsymbol{\tau}_t^{(\delta_\tau)} = \hat{\mathbf{y}}_{t|t-1} + \delta_\tau \sqrt{\operatorname{diag}(\mathbf{P}_{t|t-1})} \odot \mathbf{s}, \label{eq:threshold_mismatch}
\end{align}
where $\mathbf{s} = [+1,-1,+1]^\top$ and $\delta_\tau \in \{0,0.25,0.5,0.75,1.0\}$. 
This perturbation is scaled by the predicted measurement standard deviation and emulates systematic comparator-reference mismatch.

% As shown in Fig.~\ref{fig:threshold_mismatch_mse} and \ref{fig:threshold_mismatch_eps}, increasing $\delta_\tau$ increases the normalized mean mismatch $\epsilon_{\mu_\mathbf{z}}$ and gradually degrades the MSE.
% The degradation is smooth rather than catastrophic, indicating that adaptive thresholding should be interpreted as an approximate zero-centering mechanism rather than an exact guarantee.
% BKNet-Origin is more sensitive to this out-of-distribution threshold perturbation than BKF, whereas BKNet-Aug achieves the best MSE across the sweep.
% Since $\epsilon_{\mu_\mathbf{z}}$ is similar across the methods, the improvement of BKNet-Aug is not due to reducing the threshold mismatch itself, but due to learning a correction gain better matched to the perturbed binary observation statistics.
% Representative values are summarized in Table~\ref{tab:threshold_mismatch}.
Table~\ref{tab:threshold_mismatch} shows that increasing $\delta_\tau$ increases the mean mismatch of the quantizer input and degrades MSE smoothly rather than catastrophically.
Thus, adaptive thresholding should be interpreted as an approximate zero-centering mechanism, not as an exact guarantee.
BKNet-Origin is more sensitive to this out-of-distribution threshold perturbation than BKF, whereas BKNet-Aug achieves the best MSE at $\delta_\tau=1.0$.
Since $\epsilon_{\mu_\mathbf{z}}$ is similar across the methods at the same mismatch level, the improvement of BKNet-Aug comes from learning a correction gain better matched to the perturbed binary observation statistics, rather than from reducing the threshold mismatch itself.

% These results indicate that the adaptive-threshold mechanism is useful but approximate.
% In practical front-ends, threshold offsets, reference-DAC quantization, reference settling error, comparator noise, hysteresis, and metastability may further perturb the binary observation statistics.
% A complete treatment of these effects requires circuit-aware modeling and hardware validation, while the present experiments provide an initial algorithm-level robustness check.

\begin{table*}[t]
    \centering
    \caption{Representative cold-start and mid-sequence recovery results under state-estimation perturbations on the Lorenz attractor.}
    % {\color{blue}{
    \begin{tabular}{|c||c|c|c|c|c|c|c|c|}
    \hline
    $\beta$ & Method & MSE [dB] & Cold SI & Cold rec. & Cold err. [dB] & Mid SI & Mid rec. & Mid err. [dB] \\
    \hline\hline
    % \multirow{3}{*}{1}
    % & BKNet-Aug    & -16.24 & 0.016 & 1   & -12.04 & 0.087 & 38  &  -6.23 \\
    % & BKF          &   4.31 & 0.071 & 2   &  -2.09 & 0.168 & 241 &  20.24 \\
    % & BKNet-Origin &   4.49 & 0.081 & 5   &   1.98 & 0.221 & 244 &  20.12 \\
    % \hline
    \multirow{3}{*}{2}
    & BKNet-Aug    & -13.99 & 0.015 & 1   & -10.31 & 0.081 & 30  &  -2.92 \\
    & BKF          &   5.32 & 0.178 & 36  &  10.72 & 0.083 & 161 &  19.62 \\
    & BKNet-Origin &   5.67 & 0.306 & 142 &  21.01 & 0.086 & 208 &  18.86 \\
    \hline
    % \multirow{3}{*}{4}
    % & BKNet-Aug    &  -3.72 & 0.041 & 5   &  -4.90 & 0.022 & 88  &  13.12 \\
    % & BKF          &  11.23 & 0.358 & 94  &  23.65 & 0.131 & 183 &  26.21 \\
    % & BKNet-Origin &  10.55 & 0.287 & 173 &  23.81 & 0.141 & 141 &  25.79 \\
    % \hline
    \end{tabular}%}}
    \label{tab:cold_start_recovery}
\end{table*}

We further evaluate transient robustness when the predictor is temporarily inaccurate.
We inject a state-estimation perturbation
\begin{align}
    \Delta\mathbf{x} = \beta(\mathbf{s}_\mathbf{x}\odot\boldsymbol{\xi}),
    \qquad
    \boldsymbol{\xi} = [1,-1,1]^\top/\sqrt{3},
\end{align}
where $\mathbf{s}_\mathbf{x}$ is the component-wise standard deviation of the training states.
The perturbation is applied either at initialization, referred to as the cold-start perturbation, or at the middle of a length-2000 test sequence.
Table~\ref{tab:cold_start_recovery} reports the MSE, transient sign imbalance, transient error, and recovery time.

% The transient experiment shows that BKNet-Origin is highly sensitive to out-of-distribution tracking errors because it is trained only under nominal tracking conditions.
% BKF provides a model-based reference, but it still suffers from large transient errors and long recovery times after mid-sequence perturbations.
% In contrast, BKNet-Aug substantially reduces the MSE, transient sign imbalance, and recovery time across all tested perturbation magnitudes.
% The cold-start recovery times are generally shorter than the mid-sequence recovery times because the filter is still in a high-uncertainty regime at initialization, whereas the mid-sequence perturbation is injected after the filter has entered a confident tracking regime.
% Overall, these results show that adaptive thresholding is essential but approximate, and that gain learning can be made robust to threshold and transient perturbations through training augmentation.
The transient experiment shows that BKNet-Origin is sensitive to out-of-distribution tracking errors because it is trained only under nominal tracking conditions.
BKF provides a model-based reference, but it still suffers from large transient errors and long recovery times after mid-sequence perturbations.
In the representative $\beta=2$ case reported in Table~\ref{tab:cold_start_recovery}, BKNet-Aug achieves the lowest MSE and transient errors, and recovers within one step after cold-start perturbation and within 30 steps after mid-sequence perturbation.
In contrast, BKF and BKNet-Origin require substantially longer recovery times.
The same qualitative trend was observed for $\beta=1$ and $\beta=4$.
Overall, these results show that adaptive thresholding is useful but approximate, and that gain learning can be made more robust to threshold and transient perturbations through training augmentation.
% }}%

% \begin{table}[t]
%     \centering
%     \caption{Effect of adaptive thresholding on BKNet}
%     \begin{tabular}{|c|c|c|}
%         \hline
%         Dataset & w/o threshold update & Adaptive thresholding  \\ \hline \hline
%         Lorenz  & -14.26               & -17.31                 \\ \hline
%         NCLT    & 135.4                & 18.62                  \\ \hline
%     \end{tabular}
%     \label{tab:threshold_ablation_bknet}
% \end{table}

%-------------------------------------------------------------------
\subsection{Real-World Dynamics: Michigan NCLT Dataset} \label{section:exp_NCLT}

Finally, we evaluate the proposed methods on the Michigan NCLT dataset \cite{nclt}.
Unlike the preceding Lorenz-based experiments, exact model knowledge and branch-wise front-end control are not available for the NCLT data; therefore, we focus on partial-knowledge validation in a single-branch environment.
This dataset provides real trajectories collected from a Segway robot equipped with various sensors, including global positioning system (GPS), an odometer, and Light Detection and Ranging (LiDAR).
Since raw analog front-end waveforms are unavailable, we construct an emulated 1-bit sensing environment by re-quantizing the recorded measurements using the same thresholding rules employed in the synthetic experiments.
% {\textcolor{blue}{
Therefore, the NCLT experiment should be interpreted as real-trajectory validation of the proposed 1-bit estimation framework under emulated quantization, not as a hardware measurement of ADC energy or front-end latency.
% }}

For tracking, the state vector is defined as:
\begin{align}
    \mathbf{x}_{\tau} = (x,v_x,a_x,y,v_y,a_y)^{\top}\in\mathbb{R}^{6}.
\end{align}
For each axis, we adopt a Wiener-velocity state model with sampling interval $\Delta\tau$, where
\begin{align}
    &\mathbf{F} = \begin{pmatrix}
        1 & \Delta\tau & \frac{1}{2}\Delta\tau^2 \\
        0 & 1 & \Delta\tau \\
        0 & 0 & 1 \end{pmatrix}, \\
    &\mathbf{Q} = q^2 \begin{pmatrix}
        \frac{1}{4}\Delta\tau^4 & \frac{1}{2}\Delta\tau^3 & \frac{1}{2}\Delta\tau^2 \\
        \frac{1}{2}\Delta\tau^3 & \Delta\tau^2 & \Delta\tau \\
        \frac{1}{2}\Delta\tau^2 & \Delta\tau & 1
    \end{pmatrix}.
\end{align}
The measurement sequence consists of noisy odometer-derived velocities along each axis, so the measurement function for one axis is $\mathbf{H} = \begin{pmatrix}0 & 1 & 0\end{pmatrix}$.
Considering both axes simultaneously, the system is represented as:
\begin{align}
    &\tilde{\mathbf{F}} = \begin{pmatrix}
        \mathbf{F} & 0 \\
        0 & \mathbf{F} \end{pmatrix}, \qquad
    \tilde{\mathbf{Q}} = \begin{pmatrix}
        \mathbf{Q} & 0 \\
        0 & \mathbf{Q} \end{pmatrix}, \\
    &\tilde{\mathbf{H}} = \begin{pmatrix}
        \mathbf{H} & 0 \\
        0 & \mathbf{H} \end{pmatrix}, \qquad
    \mathbf{R} = \begin{pmatrix}
        r^2 & 0 \\
        0 & r^2 \end{pmatrix}.
\end{align}
As in the synthetic experiments, $q^2$ and $r^2$ are treated as approximate modeling parameters for the analytical filters.

For training, we use the trajectory recorded on 2012-01-22, which is sampled at $1\;\text{Hz}$ and split into 103 sequences of length $T=50$.
The training and test dates are disjoint; model selection is performed using training/validation sequences, and the length-2000 test trajectory is used only for final evaluation.
For testing, we use the trajectory from 2012-04-29, also sampled at $1\;\text{Hz}$, with a test sequence length of $T=2000$.
Since the NCLT setup provides only one measurement feature per axis, all NCLT experiments use the single-branch 1-bit front-end.

% {\color{blue}{
% % Unlike the Lorenz benchmark, the NCLT experiment contains real-data effects such as trajectory-dependent odometry errors, time-varying uncertainty, and possible model mismatch.
% % Therefore, in addition to the main performance comparison, we include a covariance-sensitivity diagnostic for BKF and a short date-dependence check for BKNet.

% Unlike the Lorenz benchmark, the NCLT experiment contains real-data effects such as trajectory-dependent odometry errors, time-varying uncertainty, and possible model mismatch.
% Therefore, in addition to the main performance comparison, we include an oracle covariance-tuning diagnostic for BKF to test whether the BKF--BKNet gap can be explained by scalar $\mathbf{Q}/\mathbf{R}$ mis-tuning.
% }}

\begin{figure}[t] 
    \centering
    \includegraphics[width=0.45\textwidth]{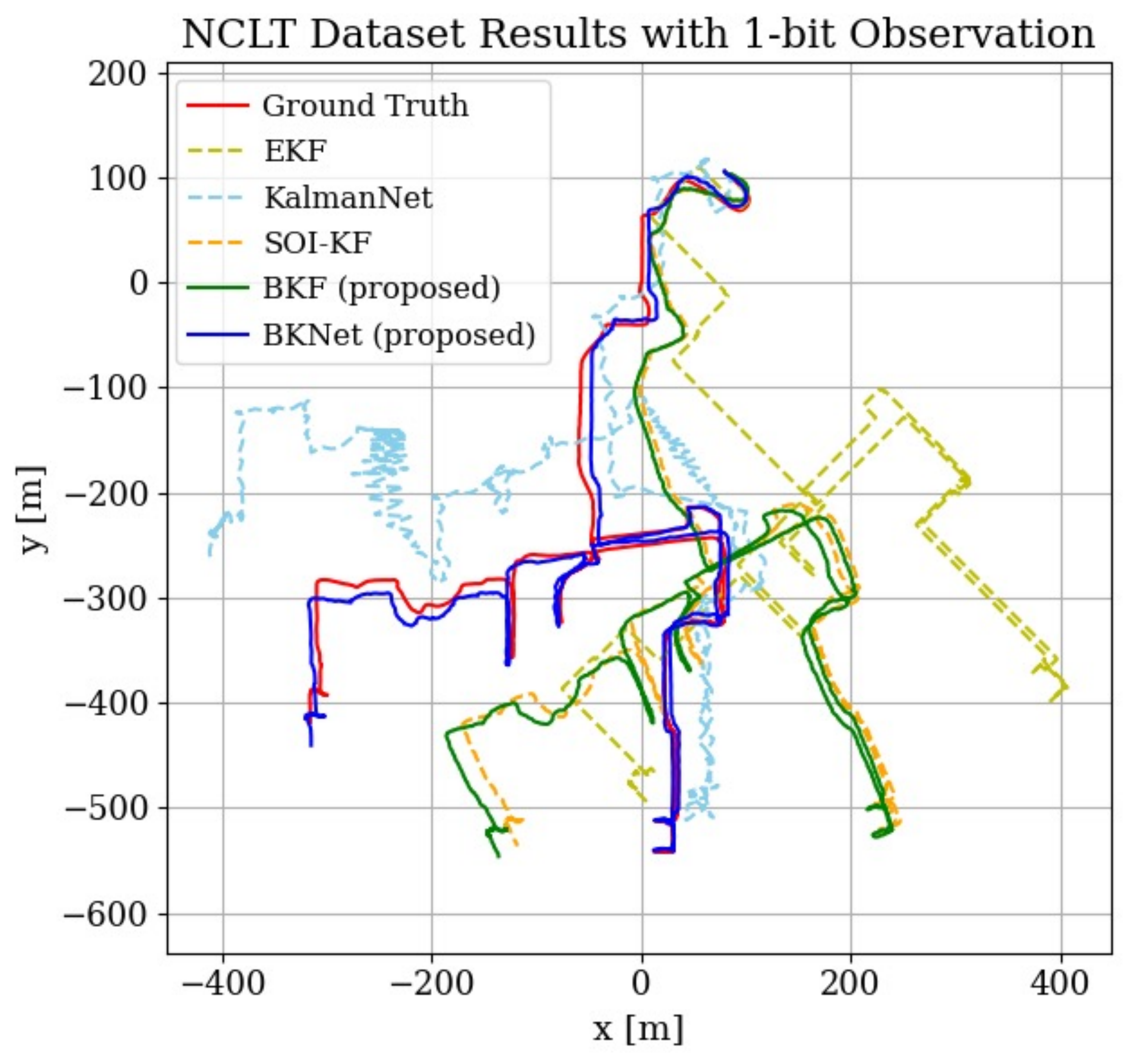}
    \vspace{-1em}
    \caption{NCLT state-estimation results under emulated 1-bit observations.}
    \label{fig:NCLT_result}
\end{figure}

Fig.~\ref{fig:NCLT_result} shows representative trajectory-level state estimation results on the NCLT dataset under 1-bit observations, illustrating that the proposed BKNet more closely tracks the ground-truth trajectory than the other baselines.
The main NCLT results are summarized in Table~\ref{tab:nclt_results}.

\begin{table}[t]
    \centering
    \caption{MSE [dB] on the Michigan NCLT Dataset}
    \label{tab:nclt_results}
    \begin{tabular}{|c||c|c|}
        \hline
        Method                & 1-bit observation    & Ideal observation  \\ \hline \hline
        EKF                   & 37.79                & 33.41              \\ \hline
        KalmanNet             & 34.67                & 19.15              \\ \hline
        Split-KalmanNet       & 40.95                & 16.58              \\ \hline
        SOI-KF                & 32.91                & --                 \\ \hline
        BKF                   & 32.58                & --                 \\ \hline
        % {\color{blue}{
        BKF, oracle $\mathbf{Q}/\mathbf{R}$-tuned%}}
                              & 29.93                & --                 \\ \hline
        BKNet                 & 18.62                & --                 \\ \hline
    \end{tabular}
    \vspace{-1em}
\end{table}

% The results demonstrate that the trends observed in the Lorenz experiments are maintained on real-world trajectories.
% While direct application of the EKF, KalmanNet, and Split-KalmanNet to 1-bit observations fails to provide reliable tracking, 
% {\textcolor{blue}{
% the quantization-aware SOI-KF improves the MSE to 32.91 dB with an inference time of 0.33 s.
% Its performance is close to BKF, which is expected because both are lightweight recursive 1-bit estimators, while SOI-KF uses a diagonal approximation of the sign-output covariance.
% BKNet still achieves the best NCLT result among the recursive 1-bit methods, indicating that learned gain adaptation is important under real-data effects such as trajectory-dependent odometry errors, time-varying uncertainty, and model mismatch.
% The exact-likelihood PF is not reported on NCLT because it is not a suitable baseline for this setting: with a six-dimensional state, single-branch 1-bit observations, long real-data rollouts, and approximate model statistics, the PF suffers from severe weight degeneracy rather than providing a meaningful near-optimal reference.
% This behavior is consistent with the well-known curse-of-dimensionality limitations of particle
% filters \cite{PF_nclt}.
% Accordingly, the likelihood-based PF is used only on the controlled Lorenz benchmark, while SOI-KF is used as the quantization-aware recursive baseline on NCLT.
Table~\ref{tab:nclt_results} shows that quantization-aware recursive filtering is necessary on NCLT.
SOI-KF improves over the direct high-resolution-estimator stress tests, and BKF gives a similar MSE, but BKNet reduces the MSE to $18.62\;\text{dB}$.
This indicates that learned gain adaptation is important under real-data effects such as trajectory-dependent odometry errors, time-varying uncertainty, and model mismatch.
The exact-likelihood PF is omitted on NCLT because the approximate real-data model and long six-dimensional rollout make it unsuitable as a near-Bayesian reference; SOI-KF is therefore used as the quantization-aware recursive baseline.
% }}

% {\color{blue}{
% To examine whether the large BKF--BKNet gap is merely due to poor scalar covariance tuning, we additionally report a test-tuned BKF result in Table~\ref{tab:NCLT_BKF_BKN}.
% In this diagnostic experiment, \(q^2\) and \(r^2\) are selected by a two-stage grid search on the test sequence.
% The selected values are $q^2 = 3 \times 10^{-4}$ and $r^2=10^{-7}$, yielding an MSE of 29.93 dB.
% This improves the nominal BKF result from $32.58\;\text{dB}$ by $2.65\;\text{dB}$, but it still remains 11.31 dB worse than BKNet.
% Therefore, the NCLT gap cannot be explained by scalar $Q/R$ tuning alone.
% The remaining gap suggests that BKNet learns gain adaptations beyond scalar covariance tuning, partially compensating for real-data effects such as time-varying odometry noise, model mismatch, and non-Gaussian disturbances.
To test whether the BKF--BKNet gap is merely due to scalar covariance mis-tuning, we perform an oracle covariance diagnostic in which $q^2$ and $r^2$ are selected by a two-stage grid search on the test sequence.
Because this tuning uses the test trajectory, the oracle $\mathbf{Q}/\mathbf{R}$-tuned BKF row is excluded from deployable method ranking and is reported only as a diagnostic upper bound on the benefit of scalar covariance retuning.
By contrast, the reported BKNet result is obtained from a checkpoint trained and selected using only the designated 2012-01-22 NCLT training and validation data; the 2012-04-29 test trajectory is not used for gradient updates, fine-tuning, or checkpoint selection.
The selected values, $q^2=3 \times10^{-4}$ and $r^2=10^{-7}$, improve BKF from $32.58$ to $29.93\;\text{dB}$, but the result remains $11.31\;\text{dB}$ worse than BKNet.
Thus, the NCLT gap cannot be explained by scalar covariance tuning alone; the learned gain appears to compensate for trajectory-dependent odometry errors, time-varying uncertainty, and non-Gaussian model mismatch.
% As an additional date-dependence check, we evaluated BKF and the same trained BKNet on another NCLT test date, 2012-08-04. BKF achieved an MSE of $34.89\;\text{dB}$ with an inference time of $0.50\;\text{s}$, whereas BKNet achieved $20.58\;\text{dB}$ with an inference time of $0.75\;\text{s}$.
% Thus, BKNet improves over BKF by $14.31\;\text{dB}$ on this additional test date, showing the same qualitative trend as the main 2012-04-29 test sequence.
% While this is not intended as an exhaustive cross-date evaluation, it suggests that the BKNet advantage observed in Table~\ref{tab:NCLT_BKF_BKN} is not unique to a single NCLT trajectory.
As an additional held-out check using the same trained BKNet checkpoint and no model selection on that date, the second NCLT test date, 2012-08-04, showed the same qualitative trend: BKNet improved over BKF by $14.31\;\text{dB}$, suggesting that the gap in Table~\ref{tab:nclt_results} is not unique to one trajectory.

% Finally, to verify that the BKNet result is not due to unstable training or test-sequence checkpoint selection, we provide the BKNet learning curves and long-rollout windowed MSE in Appendix~\ref{app:learning_curves}.
% The training and cross-validation curves show stable convergence under both Lorenz and NCLT nominal settings, and the windowed MSE confirms that BKNet remains consistently below BKF over the length-2000 NCLT rollout.
% The training and cross-validation curves show stable convergence under both Lorenz and NCLT nominal settings, and the windowed MSE confirms that BKNet remains consistently below BKF over the length-2000 NCLT rollout.
% }}

%-------------------------------------------------------------------
\section{Conclusion}
%-------------------------------------------------------------------

In this paper, we studied state estimation for resource-constrained IoT sensing systems with 1-bit sensing front-ends.
To overcome the severe nonlinearity induced by 1-bit quantization, we developed three front-end-aware recursive estimators: BKF, rBKF and BKNet.
For scenarios where the system model is fully known, the BKF integrates Bussgang linearization and adaptive thresholding into the Kalman filtering framework.
To enhance scalability in multi-branch 1-bit sensing front-ends, the rBKF reduces the dimension of the binary observation vector and provides a favorable complexity and performance tradeoff.
Built upon this reduced-order structure, BKNet learns the Bussgang gain from data, thereby improving robustness when model knowledge is partial or mismatched.

Through experiments on the Lorenz attractor with simulated 1-bit sensing and on the Michigan NCLT dataset with emulated 1-bit mobile sensing scenarios, we confirmed that the proposed methods achieve accurate state estimation even under severe quantization.
Notably, the rBKF significantly reduced inference time with only a minor performance loss in multi-branch settings, while BKNet demonstrated effective performance on real-world trajectories and under partial model knowledge.
These results suggest that front-end-aware recursive estimation represents a promising direction for low-resolution sensing and tracking in resource-constrained IoT applications.

The proposed framework opens several directions for further development, including task-adaptive projection design in rBKF, extension to few-bit quantization, and modeling of practical front-end nonidealities such as threshold mismatch and delayed updates.
Validation with real low-resolution sensing hardware is a natural next step toward demonstrating the applicability of the framework to deployed IoT systems.

%-------------------------------------------------------------------
\appendices
\section{Generalized Bussgang Linearization for Non-Zero Mean Quantizer Inputs} \label{app:nonzero_bussgang}
Let $\mathbf{u} \sim \mathcal{N}(0,\mathbf{C}_{\mathbf{u}})$ be a zero-mean Gaussian vector and let
\begin{align}
    \mathbf{z} = \boldsymbol{\gamma} + \mathbf{u},
    \qquad
    \mathbf{r} = \mathcal{Q}(\mathbf{z}),
\end{align}
where $\boldsymbol{\gamma}$ is a deterministic mean shift.
Here, $u$ denotes only the zero-mean Gaussian component of the quantizer input and is distinct from the pre-quantization measurement $\mathbf{y}_t$ used in the main text.

Define
\begin{align}
    \mathbf{a}[i] \triangleq \frac{\boldsymbol{\gamma}[i]}{\sqrt{\mathbf{C}_{\mathbf{u}}[i,i]}}.
\end{align}
Then, the mean of the 1-bit output is given element-wise by
\begin{align}
    \boldsymbol{\mu}_{\mathbf{r}}[i] \triangleq \mathbb{E}[\mathbf{r}[i]]
    = 2\Phi(\mathbf{a}[i]) - 1
    = \operatorname{erf}\! \left( \frac{\boldsymbol{\gamma}[i]}{\sqrt{2\mathbf{C}_{\mathbf{u}}[i,i]}} \right),
\end{align}
where $\Phi(\cdot)$ denotes the standard Gaussian cumulative distribution function.

The generalized Bussgang representation can be written as
\begin{align}
    \mathbf{r} = \boldsymbol{\mu}_\mathbf{r} + \bar{\mathbf{B}}\mathbf{u} + \boldsymbol{\eta},
\end{align}
where $\boldsymbol{\eta}$ is uncorrelated with $\mathbf{u}$, i.e., $\mathbb{E}[\boldsymbol{\eta} \mathbf{u}^\top] = 0$.
The diagonal generalized Bussgang matrix $\bar B$ is
\begin{align}
    \bar{\mathbf{B}}[i,i] = \sqrt{\frac{2}{\pi}} \frac{1}{\sqrt{\mathbf{C}_\mathbf{u}[i,i]}} \exp\! \left( - \frac{\boldsymbol{\gamma}[i]^2}{2\mathbf{C}_\mathbf{u}[i,i]} \right).
\end{align}

If another Gaussian random vector $\mathbf{x}$ is jointly Gaussian with $u$, then the cross-covariance between $x$ and $r$ is
\begin{align}
    \mathbf{C}_\mathbf{xr} = \mathbf{C}_\mathbf{xu} \bar{\mathbf{B}}^\top.
\end{align}
Consequently, the LMMSE estimator of $\mathbf{x}$ from the non-zero-mean 1-bit observation $\mathbf{r}$ is
\begin{align}
    \hat{\mathbf{x}} = \mathbb{E}[\mathbf{x}] + \mathbf{C}_\mathbf{xr} \mathbf{C}_\mathbf{r}^{-1}(\mathbf{r} - \boldsymbol{\mu}_\mathbf{r}),
\end{align}
where $\mathbf{C}_\mathbf{r}$ denotes the covariance matrix of the quantized output.

The main computational difficulty lies in evaluating $\mathbf{C}_\mathbf{r}$.
For the diagonal entries,
\begin{align}
    \mathbf{C}_\mathbf{r}[i,i] = 1 - \boldsymbol{\mu}_\mathbf{r}[i]^2.
\end{align}
For $i \neq j$, let
\begin{align}
    \boldsymbol{\rho}[i,j] = \frac{\mathbf{C}_\mathbf{u}[i,j]}{\sqrt{\mathbf{C}_\mathbf{u}[i,i] \mathbf{C}_\mathbf{u}[j,j]}},
\end{align}
and let $\Phi_2(\cdot,\cdot;\boldsymbol{\rho})$ denote the standard bivariate Gaussian CDF with correlation coefficient $\rho$.
Then,
\begin{equation}\begin{aligned}
    \mathbf{C}_\mathbf{r}[i,j] = &4\Phi_2(\mathbf{a}[i],\mathbf{a}[j]; \boldsymbol{\rho}[i,j])  \\
    &-2\Phi(\mathbf{a}[i]) -2\Phi(\mathbf{a}[j]) + 1 - \boldsymbol{\mu}_\mathbf{r}[i] \boldsymbol{\mu}_\mathbf{r}[j].
\end{aligned}\end{equation}
Thus, unlike the zero-mean case, the quantized-output covariance depends explicitly on the mean shift $\boldsymbol{\gamma}$ and requires evaluating mean-dependent bivariate Gaussian probabilities for all off-diagonal entries.
This can substantially increase the computational overhead in recursive filtering.

In contrast, if $\boldsymbol{\gamma} = 0$, then $\boldsymbol{\mu}_\mathbf{r} = 0$, and the generalized Bussgang matrix reduces to the zero-mean Bussgang matrix
\begin{align}
    \mathbf{B} = \sqrt{\frac{2}{\pi}} \operatorname{diag}(\mathbf{C}_\mathbf{u})^{-1/2}.
\end{align}
Furthermore, the quantized-output covariance admits the arcsine-law form
\begin{align}
    \mathbf{C}_\mathbf{r} = \frac{2}{\pi} \arcsin(\mathbf{D} \mathbf{C}_\mathbf{u} \mathbf{D}),
    \qquad
    \mathbf{D} = \operatorname{diag}(\mathbf{C}_\mathbf{u})^{-1/2},
\end{align}
where $\arcsin(\cdot)$ is applied element-wise.
This simplification is the main reason why BKF employs adaptive thresholding to make the quantizer input approximately zero-centered.
% }}

%-------------------------------------------------------------------
\section{Additional Comparison with Split-BKNet} \label{app:split_bknet}
%-------------------------------------------------------------------

%-------------------------------------------------------------------
\subsection{Construction of Split-BKNet}

To complement the main text comparison with BKNet, in this appendix we additionally evaluate Split-Bussgang-aided KalmanNet (Split-BKNet).
This variant maintains the same adaptive-thresholding front-end and reduced observation path as BKNet while replacing only the gain-learning backbone with a Split-KalmanNet-style split architecture.
That is, Split-BKNet uses the same $\mathbf{r}^*_t$ and adaptive thresholding as rBKF/BKNet, but instead of learning the Bussgang gain through a single integrated gain learner, it separately learns a state statistics representation and a measurement statistics representation.

Specifically, the state update for Split-BKNet is given by:
\begin{align}
    \hat{\mathbf{x}}_{t|t} = \hat{\mathbf{x}}_{t|t-1} + \CMcal{BG}^{\text{split}}_t (\Theta_{\mathbf{\Sigma}}, \Theta_{\mathbf{P}}, \mathbf{H}) \bigl( \mathbf{r}^*_t-\hat{\mathbf{r}}^*_{t|t-1} \bigr),
\end{align}
where $\CMcal{BG}^{\text{split}}_{t}(\Theta_{\mathbf{\Sigma}},\Theta_{\mathbf{P}}, \mathbf{H})$ denotes the split Bussgang gain obtained by separately tracking the state latent covariance and measurement latent covariance representations through  distinct recurrent blocks and then combining them.
In practice, Split-BKNet uses the same input features as BKNet, namely state-difference features and reduced 1-bit observation dynamics, but processes them separately according to the split architecture rather than feeding them into a single gain learner.

An important point is that Split-BKNet is not a core contribution of this paper, but rather a stronger learning reference.
Accordingly, we do not repeat the full derivation or detailed GRU bookkeeping for Split-BKNet.
The threshold update, reduced observation construction, loss function, and optimizer remain identical to those of BKNet.
The key difference is that the gain-learning backbone is divided into two modules following the Split-KalmanNet paradigm, and these modules are trained alternately.

%-------------------------------------------------------------------
\subsection{Additional Results}

\begin{table}[t]
    \centering
    \caption{Accuracy--complexity comparison between BKNet and Split-BKNet under the adaptive-thresholding front-end}
    \begin{tabular}{|c|c|c|c|c|}
        \hline
        Dataset & \multicolumn{2}{c|}{MSE [dB]} & \multicolumn{2}{c|}{Trainable parameters} \\ \cline{2-5}
                & BKNet  & Split-BKNet          & BKNet   & Split-BKNet \\ \hline \hline
        Lorenz  & -17.31 & -18.24               & 23,928  & 149,442 \\ \hline
        NCLT    & 18.62  & 16.73                & 107,656 & 463,496 \\ \hline
    \end{tabular}
    \label{tab:appendix_split_mse}
\end{table}

Table~\ref{tab:appendix_split_mse} summarizes the MSE performance and the number of trainable parameters of BKNet and Split-BKNet on the Lorenz attractor and the NCLT dataset.
For a fair comparison, both methods use the same reduced 1-bit observation sequence and the same adaptive thresholding rule.
Split-BKNet consistently achieves lower MSE than BKNet on both datasets.
However, this improvement comes with a substantial increase in model size: the number of trainable parameters increases from $23,928$ to $149,442$ on Lorenz and from $107,656$ to $463,496$ on NCLT.
These results indicate that Split-BKNet can serve as a stronger KalmanNet-family backbone within the proposed adaptive-thresholding framework, but its accuracy gain is obtained at the cost of significantly higher model complexity.
In contrast, BKNet remains more tightly aligned with the proposed 1-bit sensing front-end and provides a more favorable tradeoff between accuracy and complexity.
For this reason, we retain BKNet as the primary learning realization in the main text and present Split-BKNet in this appendix as an additional high-performance reference.

%-------------------------------------------------------------------
\section{BKNet Results for Identical-Noise Multi-Branch Configurations} \label{app:bknet}
%-------------------------------------------------------------------

For the sake of completeness, this appendix provides the detailed results for BKNet under identical-noise conditions. 
These results were omitted from the main text to avoid redundancy, as they exhibit the same qualitative trends as those observed for the BKF and rBKF.
The experimental setup is identical to that described in Section~\ref{section:exp_rBKF} and the BKNet experiments in the main text, where the number of parallel 1-bit comparator branches per measurement feature is varied among $1$, $8$, $64$, and $128$.

\begin{figure}[t]
    \centering
    \includegraphics[width=0.5\textwidth]{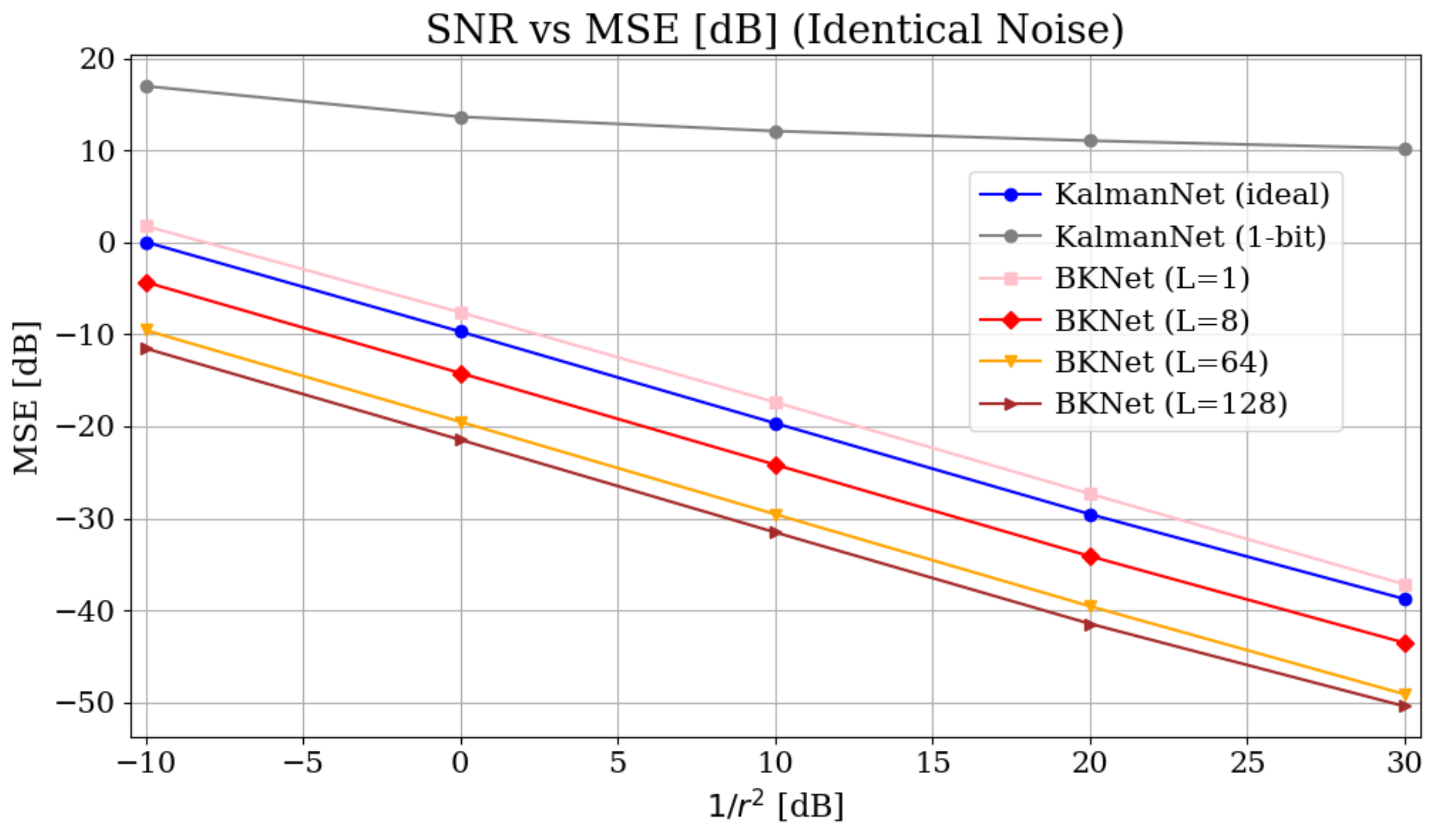}
    \vspace{-2em}
    \caption{MSE performance of BKNet under identical-noise multi-branch configurations.}
    \label{fig:BKN_identical_MSE}
\end{figure}

Fig.~\ref{fig:BKN_identical_MSE} shows the estimation performance of BKNet as a function of the number of branches and the measurement noise level under identical-noise conditions.
BKNet achieves lower MSE as the number of branches increases, confirming that a multi-branch front-end provides more useful information than the single-branch case even at the same noise level. 
In addition, the performance improves across all branch settings as the measurement noise decreases, demonstrating that BKNet can stably learn the gain even from reduced 1-bit observations.

\begin{table}[t]
    \centering
        \caption{BKNet, $\#$ of comparator branches $L$ vs Inference time [s], identical-noise}
    \begin{tabular}{|c|c|c|c|c|}
        \hline
        $L$                 & 1        & 8        & 64        & 128                \\ \hline \hline
        KalmanNet (ideal observation)          & \multicolumn{4}{c|}{1.014}    \\ \hline
        BKNet               & 1.032    & 1.032    & 1.037     & 1.049              \\ \hline
    \end{tabular}
    \label{tab:BKN_identical_interference_time}
\end{table}

Table~\ref{tab:BKN_identical_interference_time} shows the inference time under identical-noise conditions. 
The inference time of BKNet increases only slightly, remaining within the range of 1.03–1.05 seconds even as the number of branches increases.
This is because, although the multi-branch front-end produces more binary outputs, BKNet operates on a reduced observation path and therefore feeds a low-dimensional representation into the gain-learning block. 
As a result, BKNet attains improved accuracy from branch diversity in identical-noise multi-branch environments while incurring only a very small increase in computational cost.

%-------------------------------------------------------------------
% {\color{blue}{
\section{BKNet Training Behavior and Long-Rollout Stability} \label{app:learning_curves}
%-------------------------------------------------------------------

This appendix provides additional diagnostics for the training and validation behavior of BKNet under the nominal Lorenz and NCLT settings.

\begin{figure}[t]
    \centering
    \includegraphics[width=0.48\textwidth]{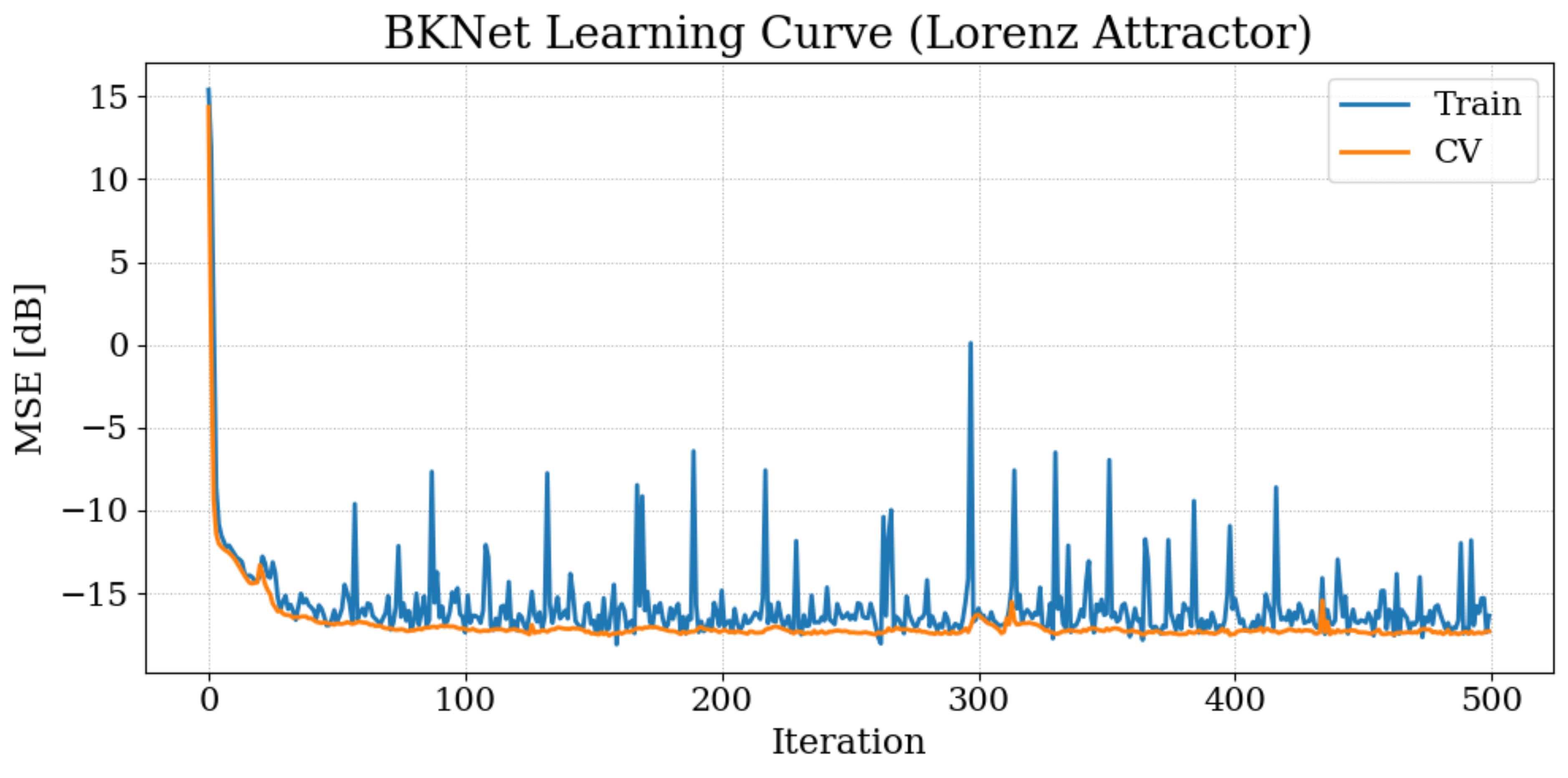}
    \includegraphics[width=0.48\textwidth]{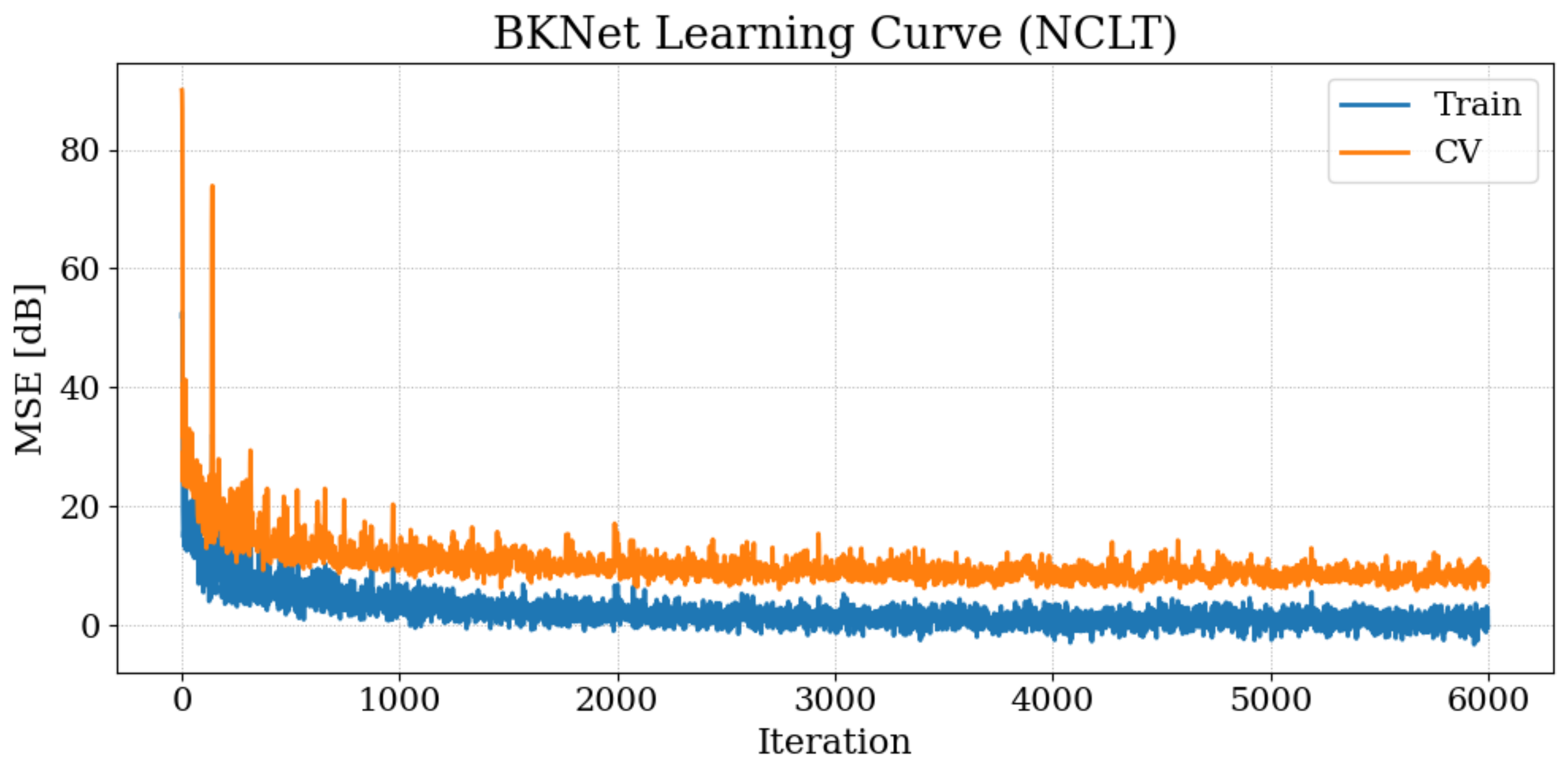}
    \caption{Training and cross-validation MSE of BKNet under the nominal Lorenz and NCLT settings. Top: Lorenz attractor. Bottom: Michigan NCLT.}
    \label{fig:learning_curves}
\end{figure}

% \begin{figure*}[!t]
%     \centering
%     \includegraphics[width=0.48\textwidth]{image_rev/learning_curve_Lorenz.pdf}
%     \hfill
%     \includegraphics[width=0.48\textwidth]{image_rev/learning_curve_NCLT.pdf}
%     \caption{Training and cross-validation MSE of BKNet under the nominal Lorenz and NCLT settings. Left: Lorenz attractor. Right: Michigan NCLT.}
%     \label{fig:learning_curves}
% \end{figure*}

Fig.~\ref{fig:learning_curves} shows the training and cross-validation MSE of BKNet.
For the Lorenz attractor, both curves decrease rapidly during the early iterations and then stabilize near the final operating region.
Although the training curve contains occasional mini-batch fluctuations under chaotic trajectory sampling, the cross-validation curve remains stable and provides no visible evidence of late-iteration overfitting.
For NCLT, the curves are noisier and exhibit a within-date validation gap, but both enter stable operating regions without late-iteration divergence.
The NCLT training and cross-validation subsets are constructed only from the 2012-01-22 trajectory, whereas the 2012-04-29 and 2012-08-04 test trajectories are not used for gradient updates or checkpoint selection.
These curves characterize the optimization behavior of BKNet; its behavior on substantially longer length-$2000$ test trajectories is evaluated in Section~\ref{section:numerical_exp}.

\bibliographystyle{IEEEtran}
\bibliography{iotj_ref.bib}

\end{document}